\documentclass{article}

\usepackage{preamble_cardiomath}
\usepackage{preamble_mine}

\title{Fast generation of 3D flow obstacles from parametric surface models:
application to cardiac valves}
\author{Bob van der Vuurst, Ji\v{r}\'i Kosinka, Crist\'obal
    Bertoglio\footnote{Corresponding author: \texttt{c.a.bertoglio@rug.nl}} \\
{\small \textit{Bernoulli Institute, University of Groningen, The Netherlands}}}
\date{\today}

\begin{document}

\maketitle

\begin{abstract}

    Due to the computationally demanding nature of fluid-structure interaction
    simulations, heart valve simulation is a complex task. A simpler alternative is to
    model the valve as a resistive flow obstacle that can be updated dynamically
    without altering the mesh, but this approach can also become computationally
    expensive for large meshes.

    In this work, we present a fast method for computing the resistive flow obstacle
    of a heart valve. The method is based on a parametric surface model of the valve,
    which is defined by a set of curves. The curves are adaptively sampled to create a
    polyline representation, which is then used to generate the surface. The surface
    is represented as a set of points, allowing for efficient distance calculations to
    determine whether mesh nodes belong to the valve surface. We introduce three
    algorithms for computing these distances: minimization, sampling, and
    triangulation. Additionally, we implement two mesh traversal strategies:
    exhaustive node iteration and recursive neighbor search. The latter significantly
    reduces the number of distance calculations by only considering neighboring nodes.
    Our pipeline is demonstrated on both a previously reported aortic valve model and
    a newly proposed mitral valve model, highlighting its flexibility and efficiency
    for rapid valve shape updates in computational simulations.

\end{abstract}

\section{Introduction}

Valvular heart disease refers to disorders affecting the heart valves, commonly due to
aging, congenital defects, or diseases like rheumatic heart disease
\cite{coffeyModernEpidemiologyHeart2016}. The main functional consequences are
stenosis (narrowing of the valve opening, restricting blood flow) and
insufficiency/regurgitation (incomplete valve closure, allowing backflow). These
conditions may occur separately or together, depending on the underlying valve alterations.

The mechanical behavior of valve tissue is highly complex and varies between subjects,
leaflets, and valves due to its intricate fiber structure
\cite{heydenMaterialModelingCardiac2015}. Experimental tests often fail to replicate
in vivo loading, as real valves experience both traction and compression, unlike the
pure traction applied ex vivo. Consequently, fully subject-specific fluid-solid
interaction (FSI) simulations remain impractical despite advances in numerical methods
\cite{astorinoFluidstructureInteractionMultibody2009,anneseSplittingSchemesLagrange2022,kaiserDesignbasedModelAortic2021,burmanMechanicallyConsistentModel2022,fernandezUnfittedMeshSemiimplicit2021}.



An alternative, but increasingly used strategy for representing valves in large-scale
cardiac (fluid-)mechanics simulations is the \textit{resistive immersed surfaces}
(RIS) method
\cite{astorinoRobustEfficientValve2012,thisAugmentedResistiveImmersed2020}. Here,
valves are modeled as internal surfaces within the mesh, where velocity is penalized
and pressure jumps are introduced via additional pressure degrees of freedom. This
allows for flexible activation or deactivation of valve configurations during simulations.

In order to completely remove the influence of the valve shape on the fluid mesh,
several authors proposed to include valves as \textit{resistive immersed implicit
surfaces} (RIIS), i.e. in terms of a resistive volumetric function
\cite{laadhariNumericalModelingHeart2016,fedelePatientspecificAorticValve2017,fumagalliImagebasedComputationalHemodynamics2020,fuchsbergerIncorporationObstaclesFluid2022,fumagalliReduced3D0DFluid2025},
which converges to the problem with a fixed obstacle when the value of the resistance
increases as proven in \cite{aguayoAnalysisObstaclesImmersed2022}. The position of the
valves can be therefore determined without defining them in the mesh \textit{a
priori}. To do so, a distance function needs to be constructed, which is then
threshold to represent the valve with a certain thickness. 

Recently in \cite{paseParametricGeometryModel2023}, a generalized parametric model of
the aortic valve was presented (i.e. by defining quantities like radii, angles etc.),
adapted to generate RIIS and therefore suitable for CFD simulations of (stenotic)
aortic valves. However, several challenges remained. Firstly, the computational cost
of generating the 3D valve obstacle surpassed the one of running the CFD simulation
itself. Secondly, though appealing, the generalization to other valves remained unaddressed.


Therefore, in this work we present and compare a variety of approaches for
considerably speeding up the conversion of parametric heart valves' representations to
flow obstacles, including an adaptive point representation of the heart valve surfaces
and three methods to calculate distances to the surface. Next, a recursive neighbor
search on the mesh nodes is also introduced that greatly reduces the number of
calculations compared to iteration over the whole mesh. Last but not least, we also
present a parametric model of the mitral valve, and assess the performance of the
presented methods. In short, the fastest of the methods is capable to generate 3D RIIS
representations of the valves with in $\mathcal{O}(10^6)$-nodes meshes in just a few
seconds, considerably surpassing the dozens of minutes in $\mathcal{O}(10^5)$-nodes
meshes reported in \cite{paseParametricGeometryModel2023}.

The remainder of this article is organized as follows. Section~\ref{sec:valve_models}
provides an overview of the aortic and mitral valve models.
Section~\ref{sec:surface_generation} details the parametric surface generation
pipeline, including curve definitions and surface construction.
Section~\ref{sec:obstacle_generation} describes the algorithms for distance
computation and mesh traversal to generate 3D flow obstacles. Performance evaluations
and results are presented throughout the relevant sections. Conclusions are given in
Section~\ref{sec:conclusions}. The detailed parametric mitral valve model is described
in  Appendix~\ref{app:mitralvalve}.

\section{Heart valve models' overview} \label{sec:valve_models}

In this section we briefly discuss the geometries of the aortic valve and the mitral
valve. The aortic valve is discussed in detail in
\cite{paseParametricGeometryModel2023} and the mitral valve is discussed in
Appendix~\ref{app:mitralvalve}. The heart valve parameters used for the figures are
listed in Table~\ref{tab:valve_params}. 

\subsection{Aortic valve}
\begin{figure}[htbp]
    \centering
    \includegraphics[width=0.8\linewidth]{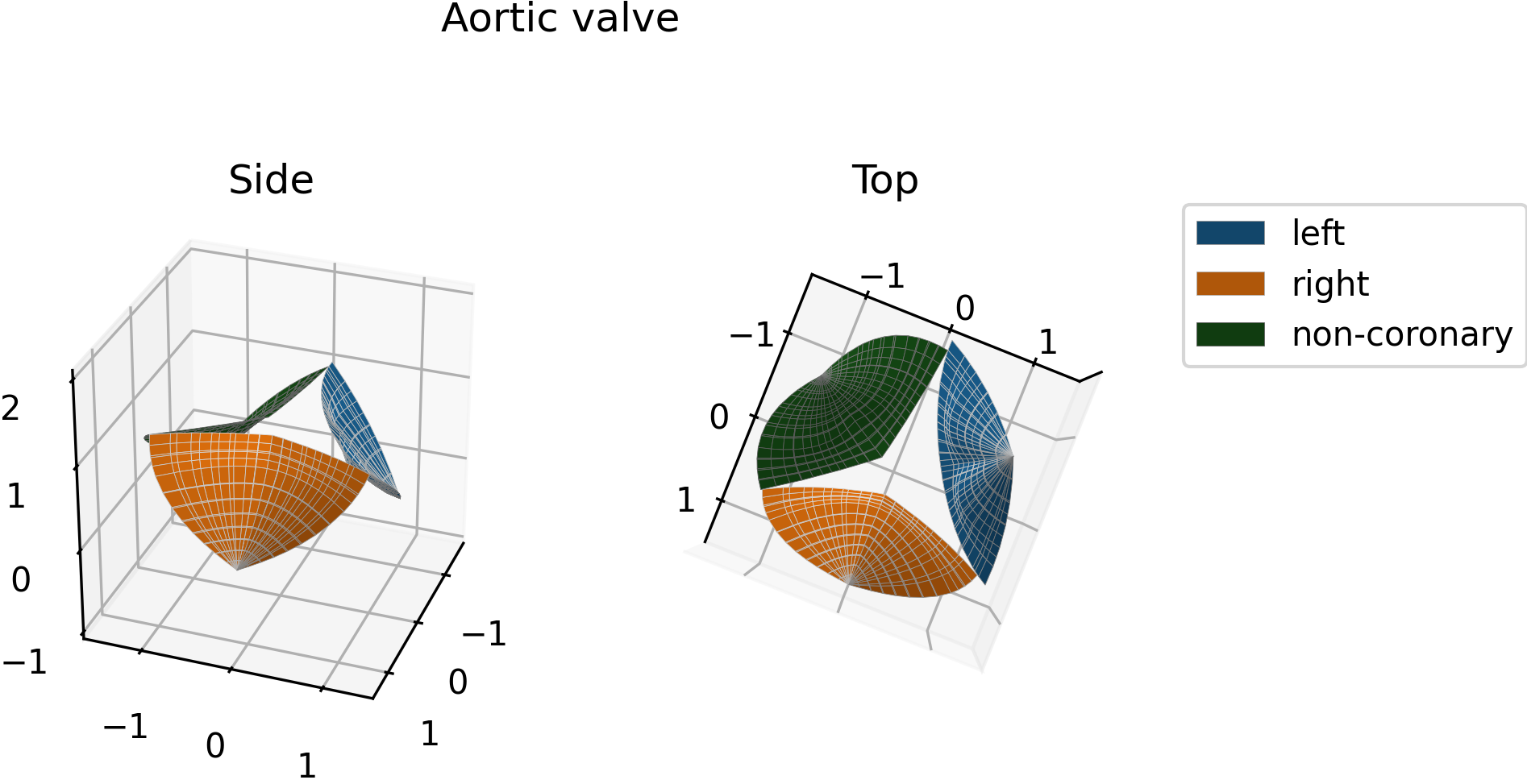}
    \caption{Overview of the cusps in the aortic valve. }
    \label{fig:aortic_valve}
\end{figure}

The aortic valve controls the blood flow between the left ventricle and the aorta. It
has three cusps, which are the left, right and con-coronary cusps. Each cusp has its
own set of parameters for its shape. This can be seen in
Figure~\ref{fig:aortic_valve}, where the left cusp has a less sharp top curve than the other two.

\begin{figure}[htbp]
    \centering
    \includegraphics[width=0.6\linewidth]{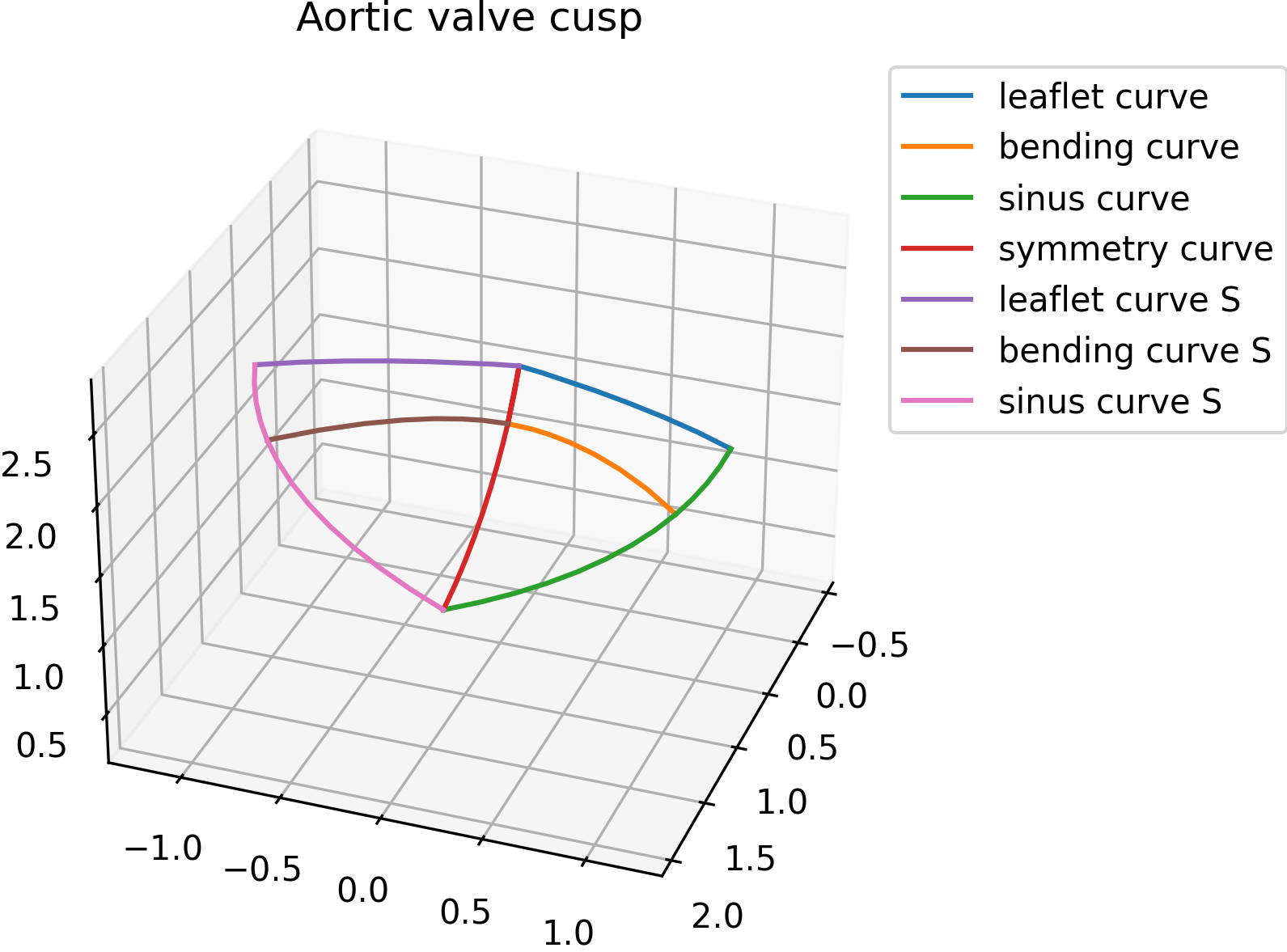}
    \caption{Curves used for defining the aortic valve cusp.}
    \label{fig:aortic_cusp}
\end{figure}

A cusp is modeled using four curves: the leaflet curve, the bending curve, the sinus
curve, and the symmetry curve. These curves are shown in Figure~\ref{fig:aortic_cusp}.
The left and right sides of the cusp are symmetric with respect to the symmetry curve.
To generate the cusp surface, it is split into top and bottom parts by the bending
curve. The top and bottom parts of the surface are then generated by bilinear
interpolation of their respective enclosing curves.

\subsection{Mitral valve}

\begin{figure}[htbp]
    \centering
    \begin{subfigure}{0.49\linewidth}
        \includegraphics[width=\linewidth]{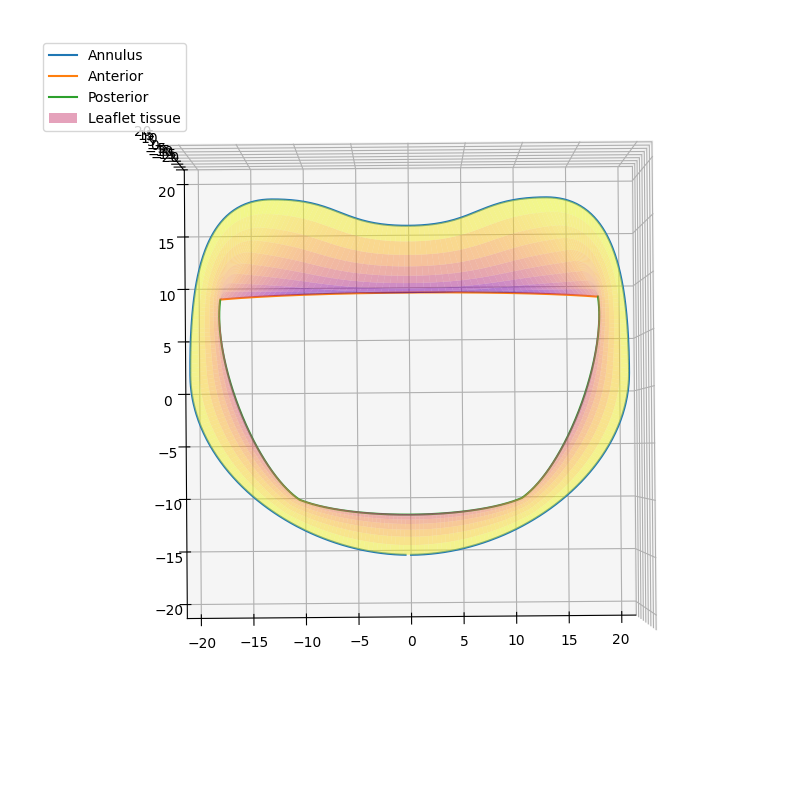}
        \subcaption{Top view}
    \end{subfigure}
    \begin{subfigure}{0.49\linewidth}
        \includegraphics[width=\linewidth]{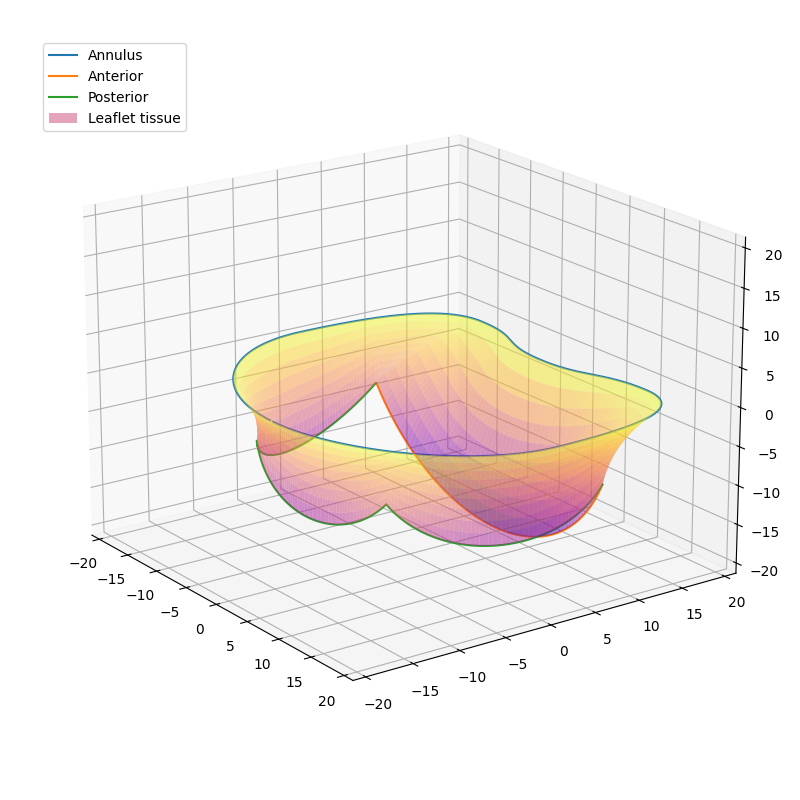}
        \subcaption{Side view}
    \end{subfigure}
    \caption{Example of mitral valve in open state. }
    \label{fig:mitral_valve}
\end{figure}

The mitral valve controls the flow of blood from the left atrium to the left
ventricle. It has two cusps: the anterior leaflet and the posterior leaflet. The
posterior leaflet has three scallops. The annulus of the mitral valve is elliptical
with an indentation at the anterior end, close to the aortic valve. The general shape
of the mitral valve is shown in Figure~\ref{fig:mitral_valve}. The mitral valve is
modeled using three curves: the annulus curve and the anterior and posterior leaflet
curves. The valve surface is generated by interpolating points on the annulus to the
anterior or posterior leaflet. This is explained in detail in Section~\ref{sec:mitral_surfaces}.

\section{Parametric surface generation} \label{sec:surface_generation}

\subsection{Overview of the pipeline}

\begin{figure}[htbp]
    \centering
    \includegraphics[width=1.0\textwidth]{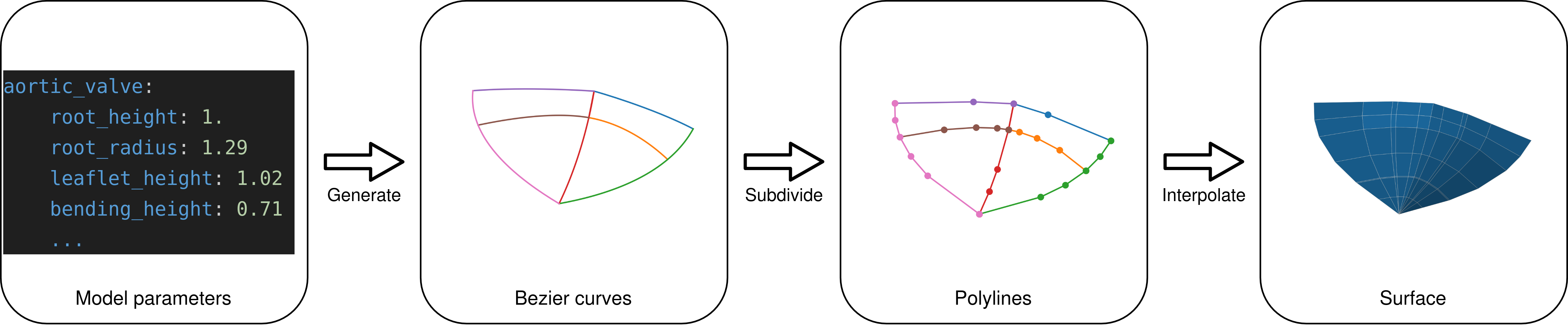}
    \caption{Pipeline of generating valve surface from parameters.}
    \label{fig:surface_pipeline}
\end{figure}

An overview of the surface generation pipeline is shown in Figure~\ref{fig:surface_pipeline}.
The pipeline starts with the definition of a number \textit{shape parameters} that
define the curves that define the heart valve surface (radius, curvature, heights,
etc). These curves are defined as Bézier curves, which are then adaptively subdivided
to get a polyline representation. The polylines are then used to generate the surface
by interpolating between them. The surface is represented as a set of points, which
can be used to calculate distances to the surface. The point representation of the
surface allows for a faster computation of distances than using the Bézier curves
directly. Finally, a mesh traversal algorithm is used to find all nodes in a
volumetric mesh that belong to the surface.

The next sections will discuss the methods used for defining the curves
(Section~\ref{sec:curves}), generating the surface (Section~\ref{sub:surface}),
calculating distances to the surface (Section~\ref{sub:distance}) and traversing the
mesh to find all nodes that belong to the surface (Section~\ref{sec:mesh_traversal}).
The performance of these methods is also evaluated in their respective sections.

\subsection{Curves}\label{sec:curves}
The heart valves are represented as surfaces generated by parametric curves. In this
section, we discuss the methods used for defining the curves. They are first defined
as Bézier curves, which are then adaptively subdivided to get a polyline
representation. The shapes of the curves are defined using control points. Two control
points define the endpoints a curve, and the other control points influence its
curvature. These control points generally do not lie on the curve itself. The
positions of these control points can be changed based on the valve model parameters.
For example, the position of a control point can depend on the height and radius of
the heart valve. The control points for the mitral valve are specified in
Section~\ref{sec:mitral_overview}.

\subsubsection{Bézier curve}\label{sub:bezier}
Bézier curves are used for defining all curves in the heart valves, because they are
simple to define with control points, but also allow for modeling complex curvatures.

For a given sequence of control points in $\mathbb{R}^3, [Q_0,\dots,Q_n ]$, the Bézier
curve of degree $n$ is defined as \cite{farin_nurbs_1999}:
\begin{equation}
    B(t) = \sum_{i=0}^n \binom{n}{i} (1-t)^{n-i} t^i Q_i ,
\end{equation}
where $\displaystyle \binom{n}{i}$ are binomial coefficients.

In this work, we use cubic ($n=3$) Bézier curves to define most heart valve curves
because they strike a balance between modeling flexibility and computational
complexity. Quadratic ($n=2$) Bézier curves are used for the simpler parts.

\begin{figure}[htbp]
    \centering
    \begin{subfigure}[t]{0.49\linewidth}
        \includegraphics[width=\linewidth]{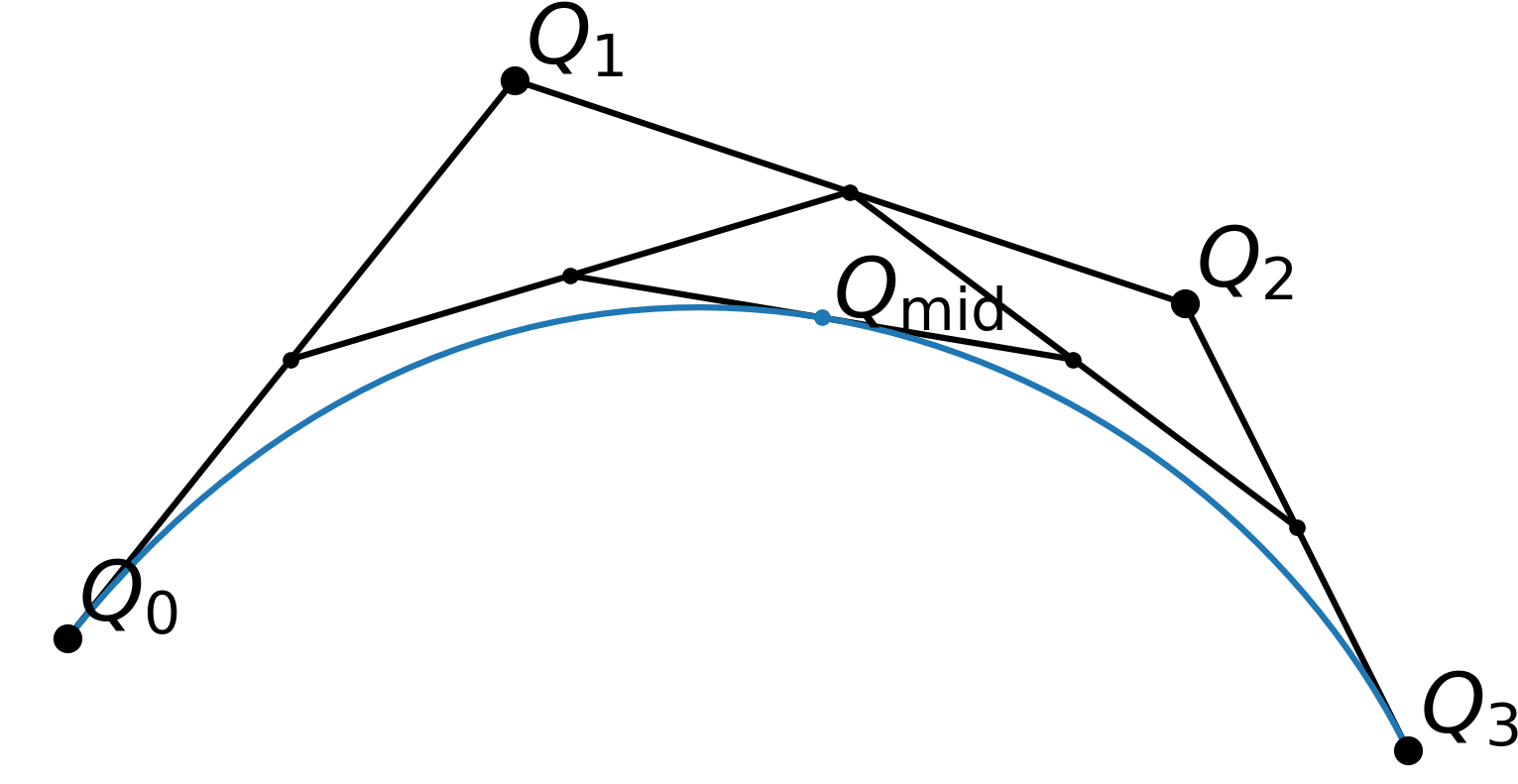}
        \subcaption{Bézier curve given by $Q_0$ to $Q_3$. }
    \end{subfigure}
    \begin{subfigure}[t]{0.49\linewidth}
        \includegraphics[width=\linewidth]{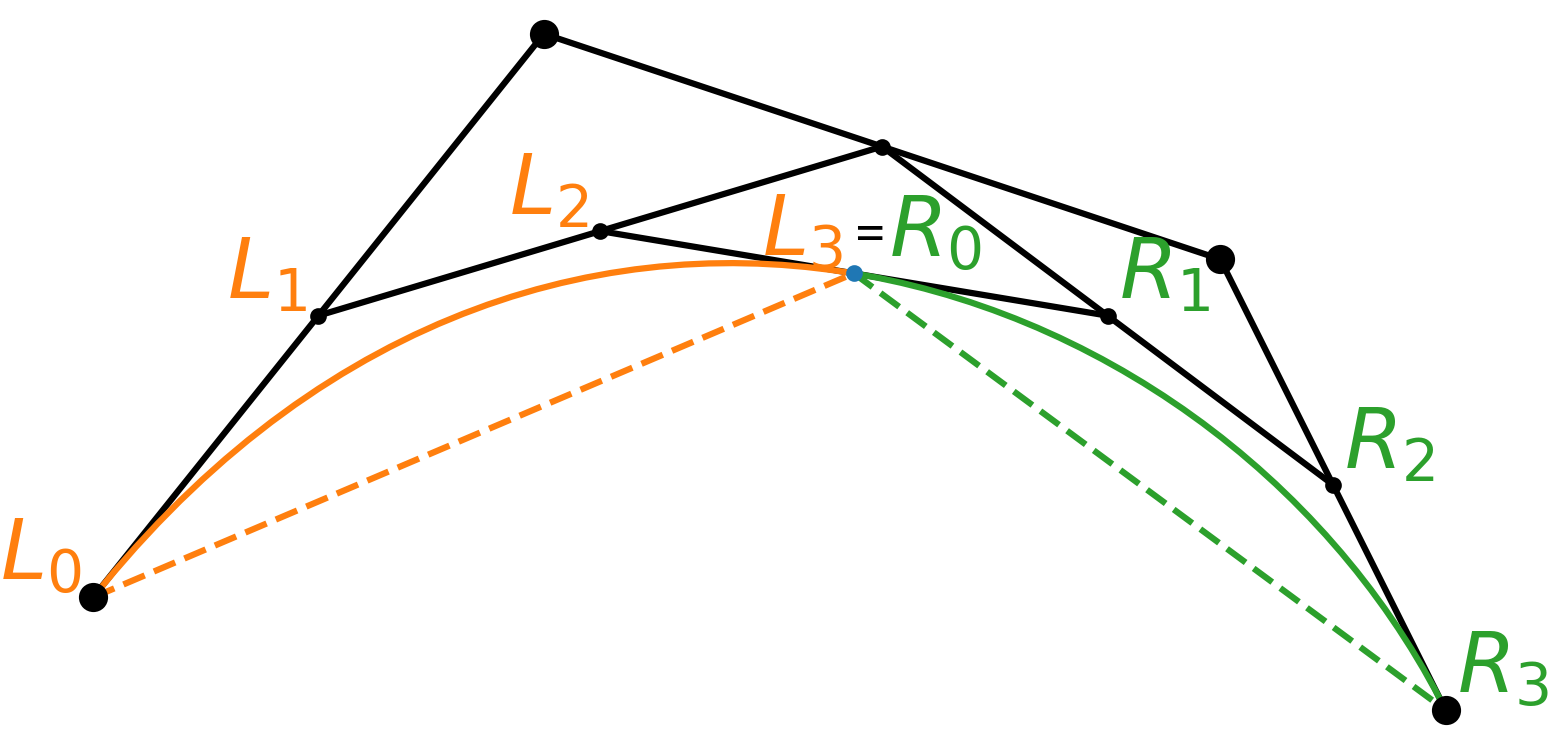}
        \subcaption{The curve split into the left arc (given by $L_0$ to $L_3$) and
        right arc (given by $R_0$ to $R_3$).} \label{fig:casteljau}
    \end{subfigure}
    \caption{Example of one step of recursive subdivision using de Casteljau algorithm.}
    \label{fig:de_casteljau}
\end{figure}

Points on Bézier curves are sampled using recursive de Casteljau subdivision (see
Algorithm~\ref{alg:casteljau} and Figure~\ref{fig:de_casteljau}), in which a Bézier
curve is split into two Bézier curves of the same order \cite{farin_nurbs_1999}. This
happens recursively until they are approximately flat when $\kappa < \phi$, given a
flatness threshold $\phi$ and flatness estimation $\kappa$:

\begin{equation}
    \kappa([Q_0, \cdots, Q_n]) = \max_{i \in [1, \cdots, n-1]} \left [ \frac{\| (Q_i -
    Q_0) \times (Q_i - Q_n) \|}{\|Q_n - Q_0 \|}  \right ].
\end{equation}
This provides adaptive sampling of Bézier curves, with areas of higher curvature
having a greater number of points. The threshold $\phi$ can be set based on mesh
density. These points can then be used to construct a polyline, as described in
Section~\ref{sub:polyline}.

\begin{algorithm}[H]
    \caption{Recursive de Casteljau subdivision for quadratic and cubic Bézier curves.}
    \label{alg:casteljau}

    \begin{algorithmic}
        \Require{Bézier $B(t)$ with order $n \in \{2,3\}$, control points $[Q_0, \cdots, Q_n]$}
        \Ensure{Sample points $[P]$}
        \Function{CasteljauSubdivision}{$n, [Q_0, \cdots, Q_n]$}
        \If{$\kappa([Q_0, \cdots, Q_n]) \leq \phi$}
        \State
        \Return $[ Q_0, Q_{n} ] $
        \Else
        \State $Q_\text{mid} \gets B(0.5)$
        \If{$n = 2$}
        \State $[L_0, L_1, L_2 ] \gets \displaystyle \left[  Q_0, \frac{Q_0 + Q_1}{2},
        Q_\text{mid} \right]$
        \State $[R_0, R_1, R_2] \gets \displaystyle  \left[Q_\text{mid}, \frac{Q_1 +
        Q_2}{2}, Q_2 \right]$
        \ElsIf{$n = 3$}
        \State $[L_0, L_1, L_2, L_3] \gets \displaystyle  \left[Q_0, \frac{Q_0 +
        Q_1}{2}, \frac{Q_0 + 2 Q_1 + Q_2}{4}, Q_\text{mid} \right]$
        \State $[R_0, R_1, R_2, R_3] \gets \displaystyle  \left[Q_\text{mid},
        \frac{Q_1 + 2 Q_2 + Q_3}{4}, \frac{Q_2 + Q_3}{2}, Q_3 \right]$
        \EndIf
        \EndIf
        \State $P_\text{left} \gets \text{CasteljauSubdivision}(n, [L_0, \cdots, L_n])$
        \State $P_\text{right} \gets \text{CasteljauSubdivision}(n, [R_0, \cdots, R_n])$
        \State \Return $[P_\text{left}, P_\text{right}]$
        \EndFunction
    \end{algorithmic}
\end{algorithm}

\subsubsection{Polyline}\label{sub:polyline}
Using recursive de Casteljau subdivision, geometry-aware polyline representations of
Bézier curves can be constructed. The polyline consists of all line segments between
each adjacent pair of endpoints, as exemplified in Figure \ref{fig:casteljau} by the
dashed polyline.
\begin{figure}[htbp]
    \centering
    \includegraphics[width=0.49\linewidth]{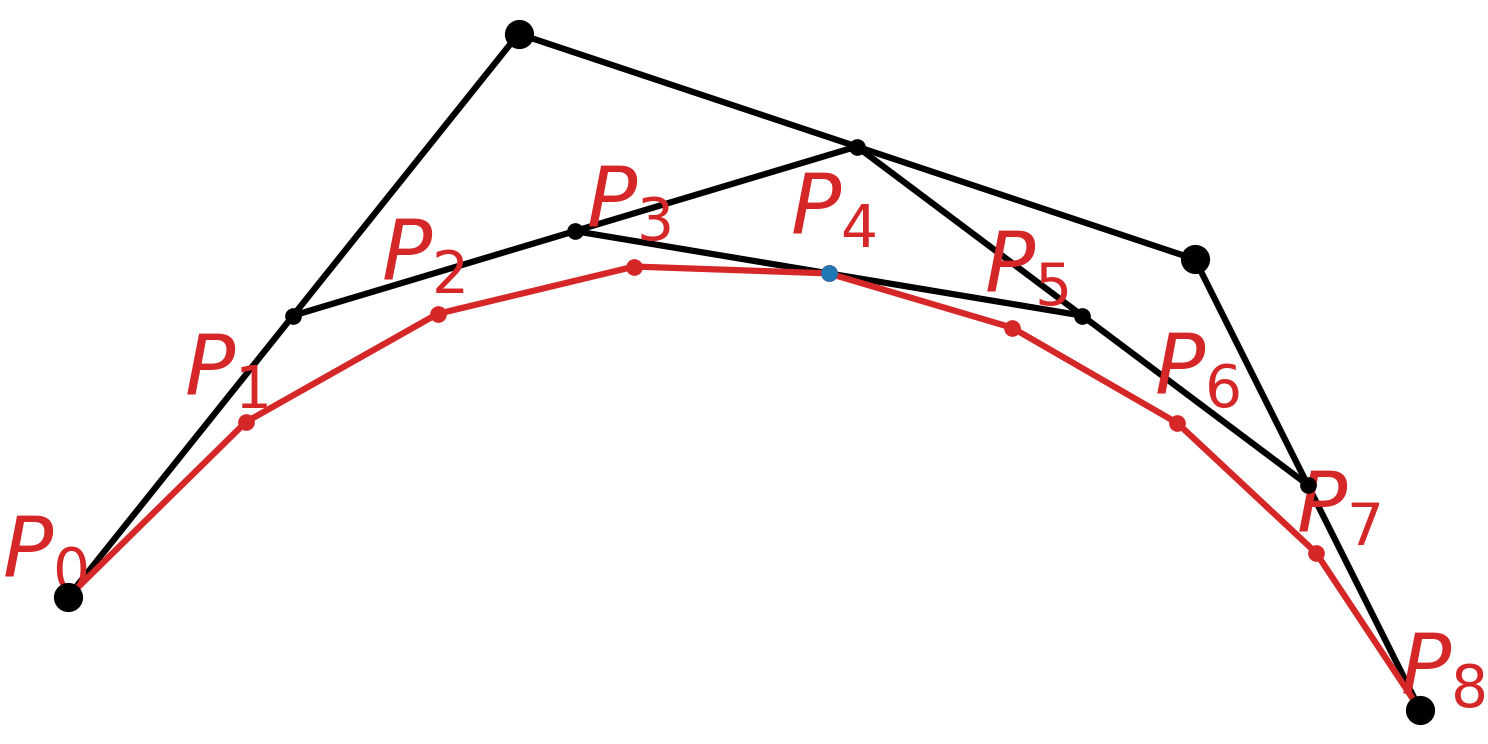}
    \caption{Polyline after recursive subdivision shown in Figure \ref{fig:casteljau}.}
    \label{fig:polylines}
\end{figure}

Similarly to Bézier curves, a point along the polyline curve can be represented using
a value $t \in [0,1]$. However, different from Bézier curves, we define $t$ to
represent the Euclidean length along the polyline. For example, $t=0.5$ is exactly at
the middle of a polyline. For each vertex $P_i$ of the polyline, its $t_i$ is defined
using the accumulated length of the curve:
\begin{equation}
    t_0 = 0, \quad t_i =    \frac{\sum_{j=1}^{i} \|P_j - P_{j-1}\| }{\sum_{k=1}^{m
    } \|P_k - P_{k-1}\|}, i>0,
\end{equation}
where the polyline has $m+1$ vertices; see Figure~\ref{fig:polylines}.

Using these values $t_i$, an arbitrary point along the polyline $C(t)$ can be found.
The first step for this is to find the minimum index $l$ where $t_l > t$ holds:
\begin{align}
    l &= \argmin_i \{t_i: t_i > t \}.
\end{align}
Using $l$, we then linearly interpolate between $P_{l-1}$ and $P_l$ to get $C(t)$:
\begin{align}
    d_l(t) &= \frac{t - t_{l-1}}{t_l - t_{l-1}} ,    \\
    C(t) &= (1-d_l(t)) P_{l-1} + d_l(t) P_l .
\end{align}
This definition of $C(t)$ gives the property that $t$ is defined as a ratio of the
length along the curve. This allows for precise placement of points along the polyline
approximation of the curve, given some curve length distance. This property is used
for the placement of some control points in the mitral valve, such as the commissure
points (Section~\ref{sec:mitral_leaflets}).

\subsection{Surface structure}\label{sub:surface}

\begin{figure}[htbp]
    \begin{subfigure}{0.32\linewidth}
        \includegraphics[width=\linewidth]{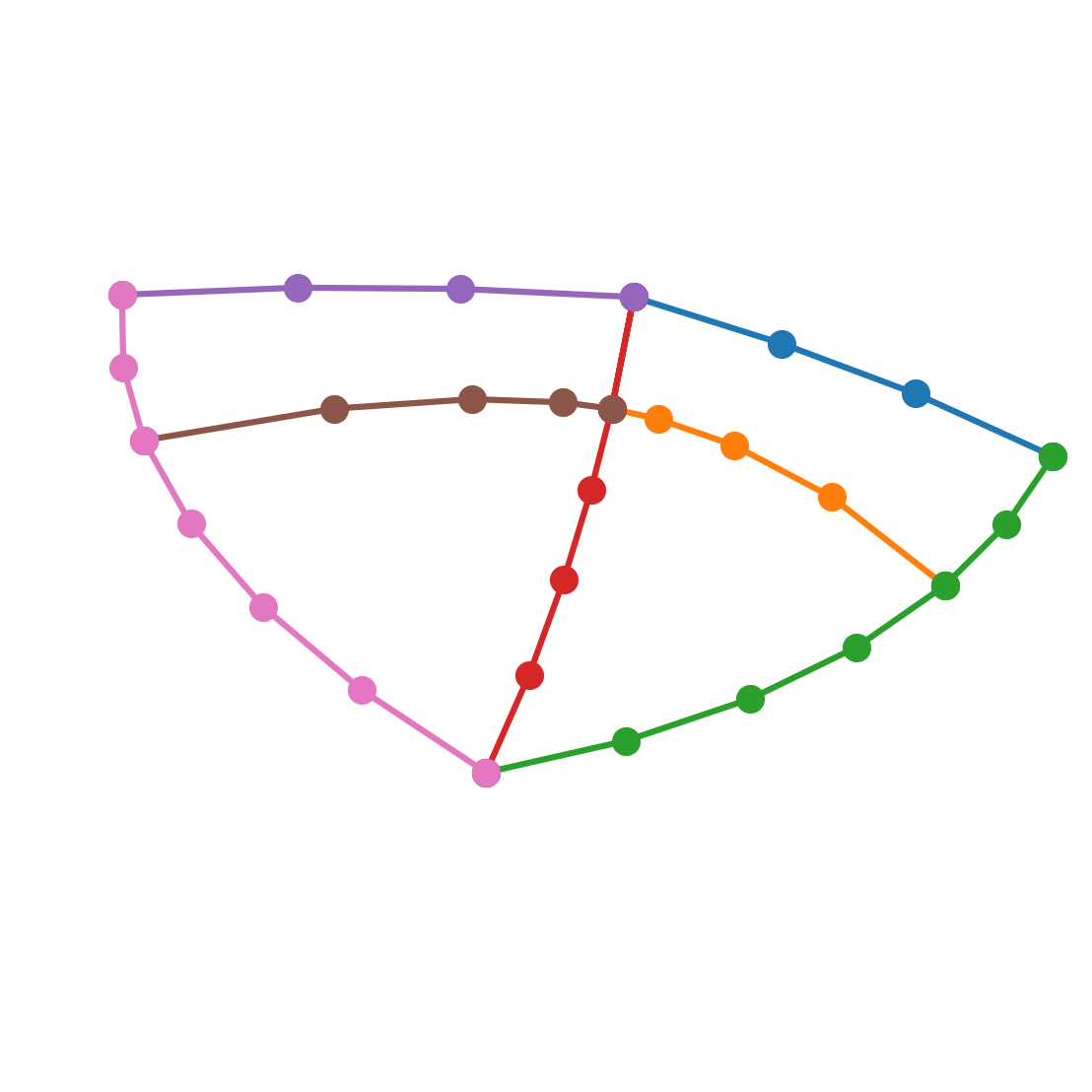}
        \subcaption{Polyline representation. }
    \end{subfigure}
    \begin{subfigure}{0.32\linewidth}
        \includegraphics[width=\linewidth]{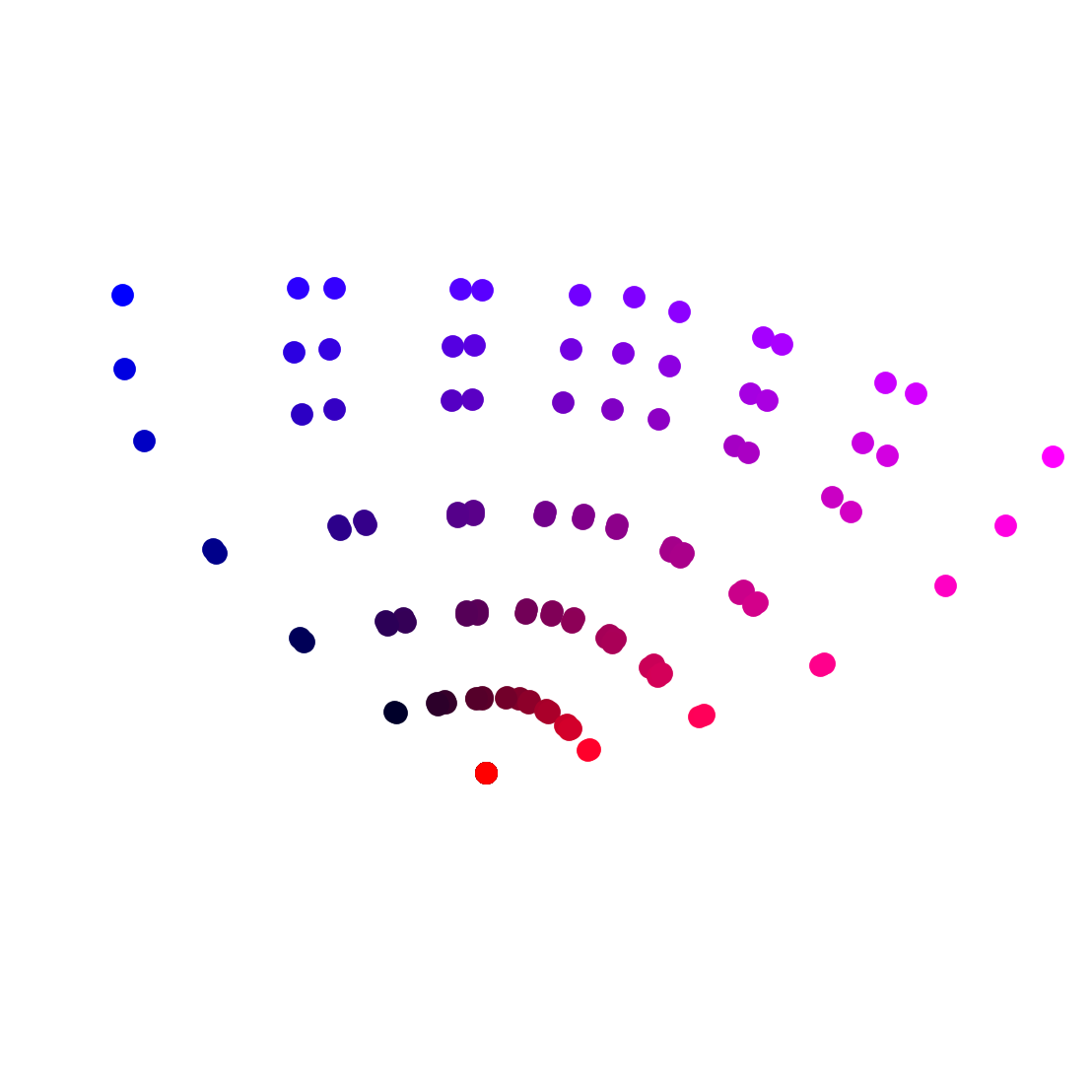}
        \subcaption{Interpolated points. }
    \end{subfigure}
    \begin{subfigure}{0.32\linewidth}
        \includegraphics[width=\linewidth]{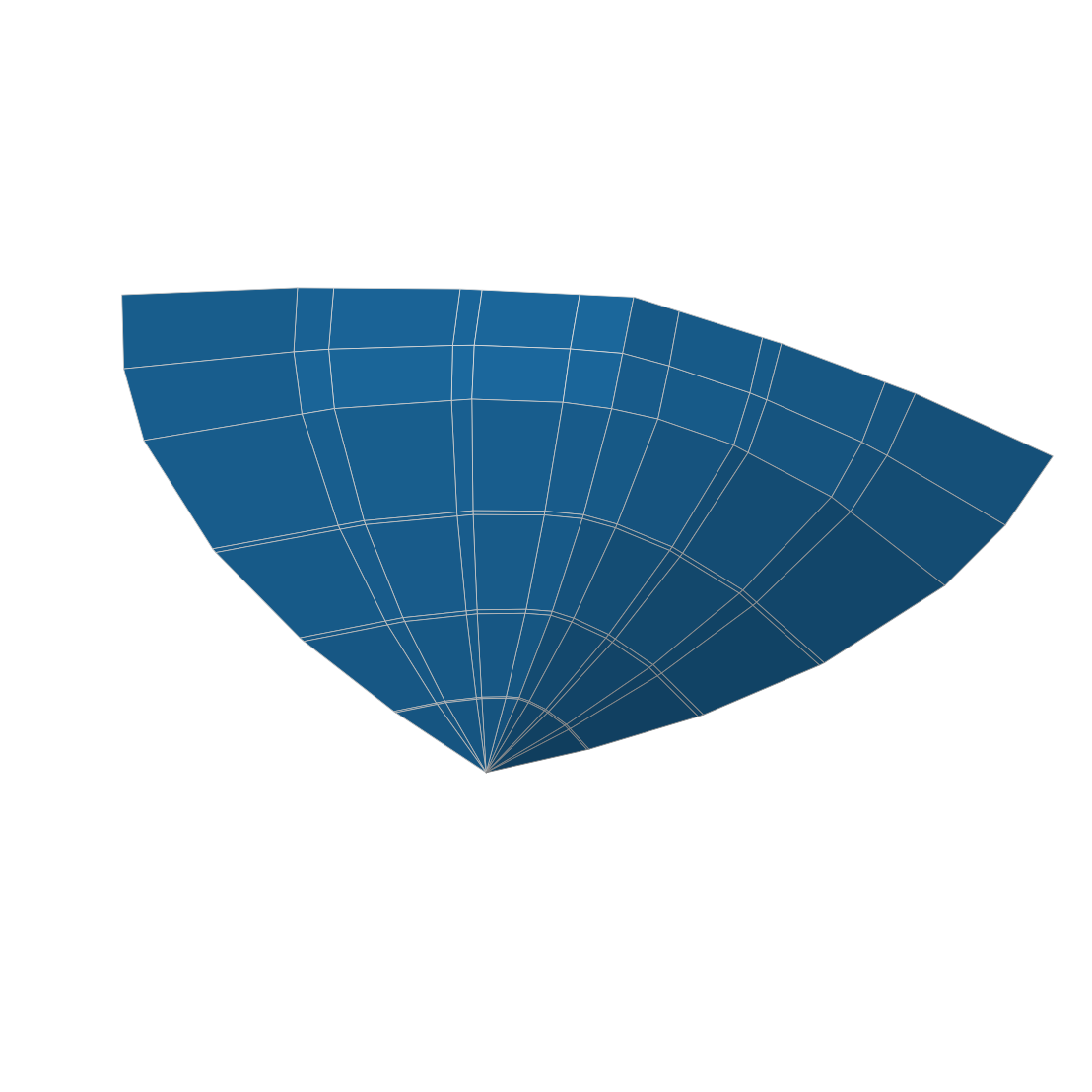}
        \subcaption{Quad representation. }
    \end{subfigure}
    \caption{Aortic valve cusp surface representations. }
    \label{fig:surface_quads}
\end{figure}

Similar to polylines, surfaces are defined using points $Q_{0,0}, Q_{0,1}, \dots,
Q_{n_u, n_v}$, which are found by bilinear interpolation of polyline curves. Each
polyline is used for interpolation of either the $u$ or $v$ direction. For example,
the aortic valve model's symmetry and sinus curves are used for $v$ interpolation. For
each direction, $u_i$ and $v_j$ are derived from the values $t_i$ of each polyline.
Here all unique values of $t_i$ for every polyline in its direction are combined into
a single list of values for $u_i$ or $v_i$. For example, two polylines with $t_i$
values $[0, 0.3, 0.6, 1]$ and $[0, 0.5, 0.6, 1]$ would result in the surface $u_i$
values $[0, 0.3, 0.5, 0.6, 1]$. The surface points are then obtained by interpolation
at each coordinate $u_i, v_i$. This ensures that no details are lost. Using the
interpolated points, the surface is represented as a quad-mesh, where every four
adjacent points form a quad. An example of a quad-mesh representation of a surface is
shown in Figure~\ref{fig:surface_quads}.

A point on the surface $S(u,v)$ with $u,v \in [0,1]$ can be defined by first finding
the quad that contains $S(u,v)$, where the indices $o, p$ are the lowest indices of
$u_i, v_j$ greater than $u, v$ respectively:
\begin{align}
    o =& \argmin_i \{u_i: u_i > u\} ,\\
    p =& \argmin_i \{v_j: v_j > v\}    .
\end{align}
The indices $o, p$ are then used for bilinear interpolation to find $S(u,v)$:
\begin{align}
    d_o(u) =& \frac{u - u_o}{u_o - u_{o-1}}         ,\\
    d_p(v) =& \frac{v - v_p}{v_p - v_{p-1}}
\end{align}
\begin{equation}
    S(u, v) = (1- d_u) (1 - d_v) Q_{{o-1}, {p-1}} + d_u (1 - d_v) Q_{o, {p-1}} + (1 -
    d_u) d_v Q_{{o-1}, p} + d_u d_v Q_{o, p} .
\end{equation}

\section{3D obstacle generation} \label{sec:obstacle_generation}

To use the surface in a volumetric simulation mesh, we need to find all nodes in the
volumetric mesh that belong to the surface, as it is done in the RIIS approach \cite{
laadhariNumericalModelingHeart2016,fedelePatientspecificAorticValve2017,fumagalliImagebasedComputationalHemodynamics2020,fuchsbergerIncorporationObstaclesFluid2022,fumagalliReduced3D0DFluid2025,paseParametricGeometryModel2023}.
This is done by calculating the distance between each node and the surface, and
checking whether it is below a threshold. The distance calculation is discussed in
Section~\ref{sub:distance}. The mesh traversal algorithm is discussed in
Section~\ref{sec:mesh_traversal}.

\subsection{Distance-to-surface computation} \label{sub:distance}

To check whether a point in the mesh belongs to the valve surface, represented with a
certain thickness $2 d_{\min}$, we need to compute the distance $D(P, S)$ between a
point $P$ and a surface $S$. We make use of three distance algorithms: minimization,
sampling and triangulation. 

\subsubsection{Algorithms}

\paragraph{Minimization}
The minimization method finds the minimum distance $D(P, S)$ using Powell's algorithm
\cite{Powell_1964}, which iteratively searches for the minimum of a cost function
without calculating derivatives. Powell's algorithm is used to find the parameters
$u,v$ that give the minimum value of $D(P, S(u,v)) = \| S(u,v) - P \|^2$. The search
is initialized with  $(u,v)=(0.5,0.5)$ and stopped when the found     minimum value is
within a given tolerance. The tolerance is calculated as a proportion  based on the
heart valve's leaflet thickness. 

\paragraph{Sampling}
The distance $D(P, S)$ can also be approximated by sampling points $S(u, v)$ on the
surface, calculating the distances and returning the point giving the minimum distance.
The points are evenly sampled along the surface $u$ and $v$ values with $s_u, s_v$
samples respectively. The number of samples affects the accuracy of the computed
distance, where a high number of samples gives a lower and more accurate minimum
distance. To balance the computational cost and accuracy, the number of samples for
$u$ and $v$ is calculated dynamically based on the width and height of the surface and
the required precision.

\paragraph{Triangulation}
Alternatively to sampling, the distance to the surface can be computed using
triangulation. Similar to the de Casteljau algorithm for polylines, surface quads can
be recursively subdivided into four smaller quads until they are approximately flat.
Then, they can each be split into two triangles. For readability a quad is redefined
using points $U_0, U_1, U_2, U_3 = Q_{i, j}, Q_{i+1, j}, Q_{i+1,j+1}, Q_{i, j+1}$ respectively.

The flatness calculation $\kappa_{\text{quad}}$ for a quad is defined as the distance
from the midpoint $U_{\text{mid}}$ of the quad to the middle of the diagonal crease
when split by two triangles:

\begin{align}
    U_{\text{mid}} =& \frac{U_0 + U_1 + U_2 + U_3}{4}, \\
    U_{\text{crease}} =& \frac{U_0 + U_2}{2}, \\
    \kappa_{\text{quad}} =& \| U_{\text{crease}} - U_{\text{quad}} \| .
\end{align}

Using the flatness calculation and a threshold $\phi_{\text{quad}}$, the quad is
recursively subdivided and then split into triangles. The algorithm for recursive quad
subdivision is shown in Algorithm~\ref{alg:triangulation}. The threshold
$\phi_{\text{quad}}$ is set by multiplying a the minimum distance by some relative
error tolerance.

\begin{algorithm}[H]
    \caption{Recursive quad subdivision for triangulation}
    \label{alg:triangulation}

    \begin{algorithmic}
        \Require{Quad defined by points $U_0, U_1, U_2, U_3$ }
        \Ensure{List of triangles $[T_0, \cdots, T_k]$}
        \Function{QuadSubdivision}{$U_0, U_1, U_2, U_3$}
        \If{$\kappa_{\text{quad}}(U_0, U_1, U_2, U_3) < \phi_{\text{quad}} $}
        \State \Return $ \displaystyle \left \{
            \begin{aligned}
                &\text{triangle}(U_0, U_1, U_2), \\
                &\text{triangle}(U_0, U_2, U_3)
        \end{aligned} \right \} $
        \Else
        \State $U_{\text{mid}}         \gets \displaystyle \frac{U_0 + U_1 + U_2 + U_3}{4}$
        \State $U_{\text{bottom}}     \gets \displaystyle \frac{U_0 + U_1}{2}$
        \State $U_{\text{right}}     \gets \displaystyle \frac{U_1 + U_2}{2}$
        \State $U_{\text{top}}         \gets \displaystyle \frac{U_2 + U_3}{2}$
        \State $U_{\text{left}}     \gets \displaystyle \frac{U_0 + U_3}{2}$
        \State \Return $ \displaystyle \left \{
            \begin{aligned}
                &\text{QuadSubdivision}(U_0, U_{\text{bottom}}, U_{\text{mid}},
                U_{\text{left}}), \\
                &\text{QuadSubdivision}(U_{\text{bottom}}, U_1, U_{\text{right}},
                U_{\text{mid}}), \\
                &\text{QuadSubdivision}(U_{\text{mid}}, U_{\text{right}}, U_2,
                U_{\text{top}}), \\
                &\text{QuadSubdivision}(U_{\text{left}}, U_{\text{mid}}, U_{\text{top}}, U_3)
        \end{aligned} \right \}  $
        \EndIf
        \EndFunction
    \end{algorithmic}
\end{algorithm}

The distance to all triangles can be computed directly, where the minimum is returned.
The main advantage of this approach is that it directly uses the quad mesh's geometry
instead of relying on samples. Our algorithm used for calculating the distance to
triangles is based on an implementation in the Renderkit Embree raytracing library,
which is optimized for performance \cite{renderkitembree_2025}.

\subsubsection{Performance evaluation}\label{sub:distance_bench}

\begin{figure}[htbp]
    \centering
    \includegraphics[width=0.6\linewidth]{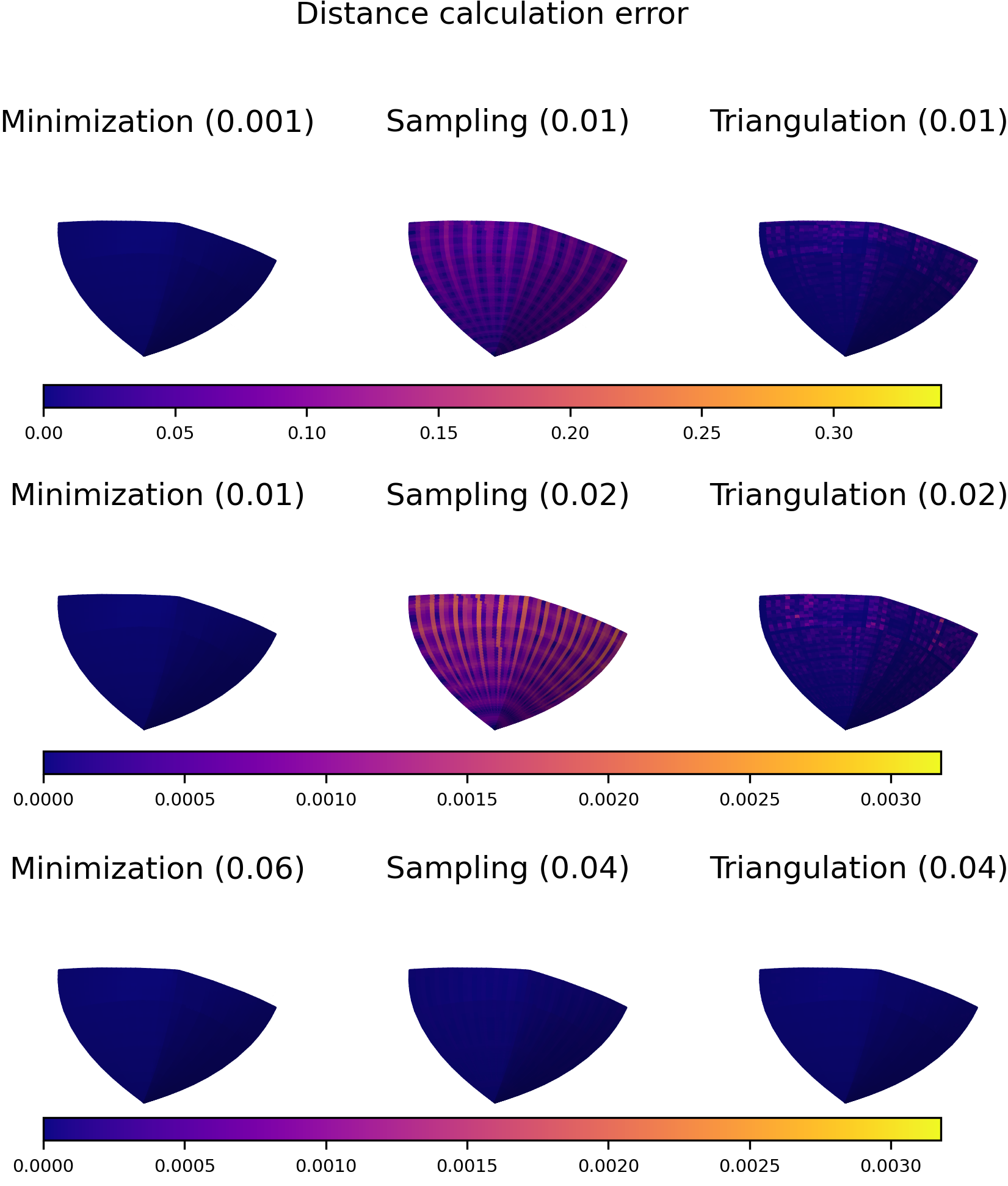}
    \caption{Errors in distance calculation plotted on aortic cusp surfaces for all
    three distance functions.}
    \label{fig:distance_acc}
\end{figure}

\begin{figure}[htbp]
    \begin{subfigure}[t]{0.48\linewidth}
        \includegraphics[width=\linewidth]{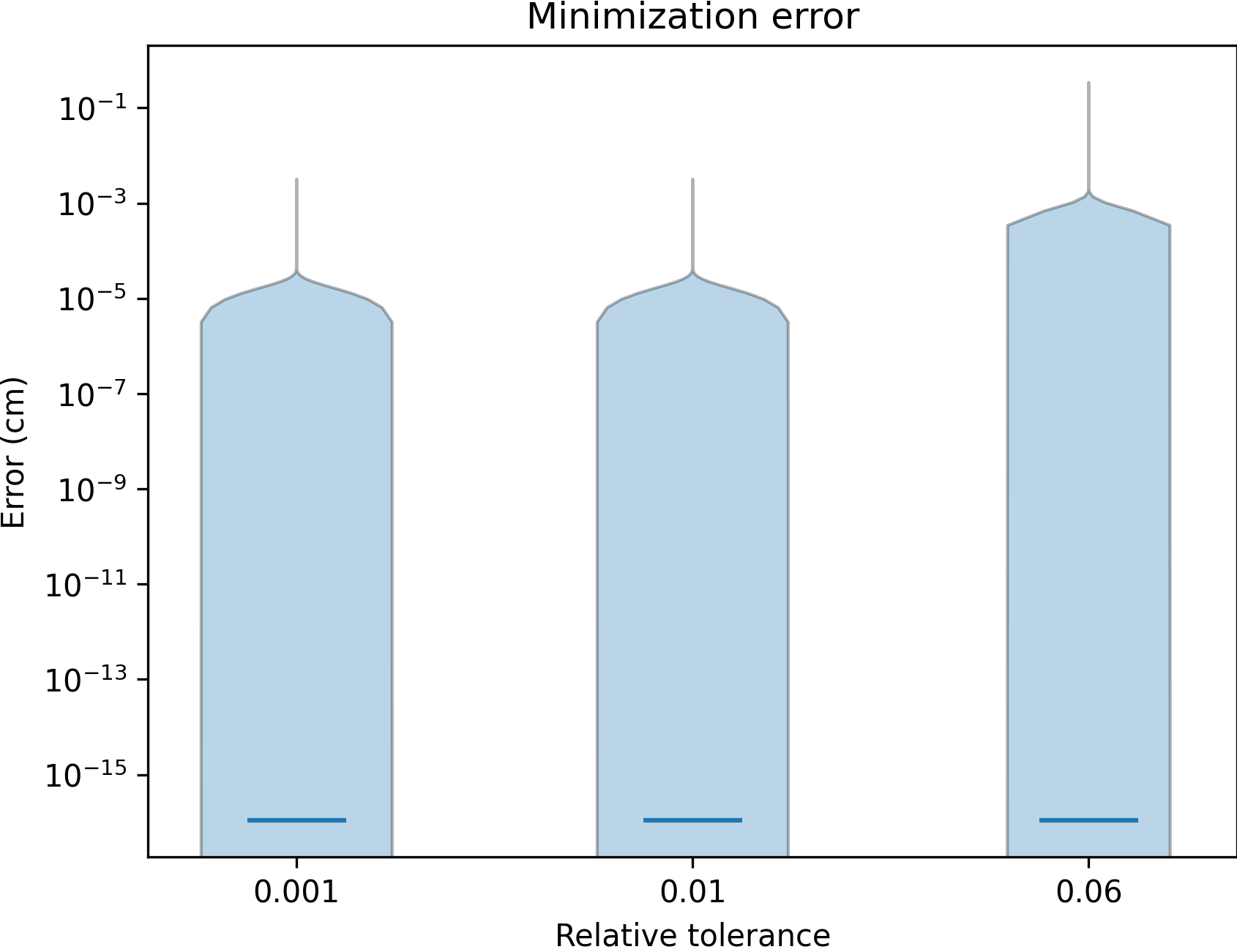}
        \subcaption{Minimization accuracy. Note the log scaling. }
    \end{subfigure}
    \hfill
    \begin{subfigure}[t]{0.48\linewidth}
        \includegraphics[width=\linewidth]{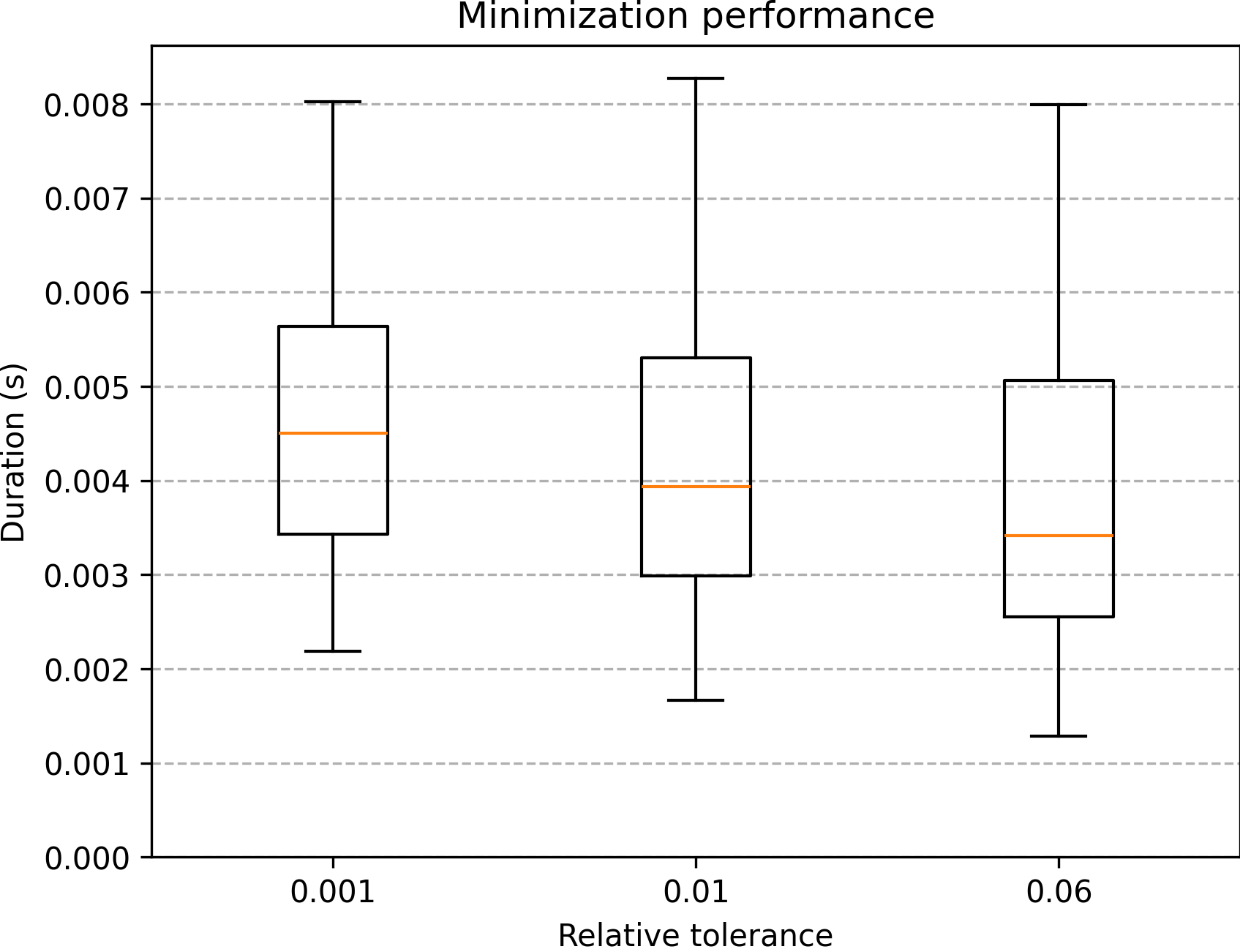}
        \subcaption{Minimization performance. }
    \end{subfigure}
    \\[\baselineskip]
    \begin{subfigure}[t]{0.48\linewidth}
        \includegraphics[width=\linewidth]{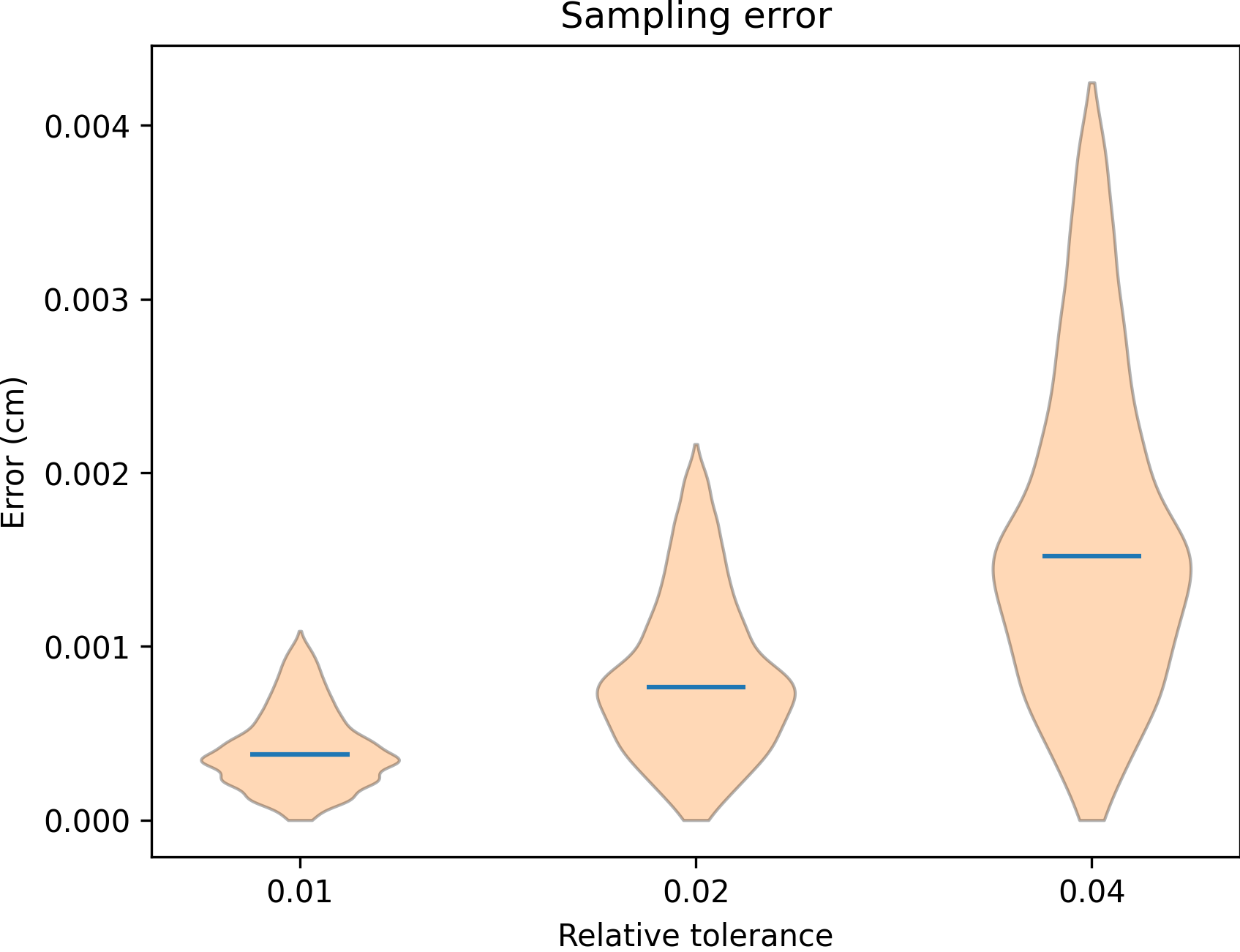}
        \subcaption{Sampling accuracy. }
    \end{subfigure}
    \hfill
    \begin{subfigure}[t]{0.48\linewidth}
        \includegraphics[width=\linewidth]{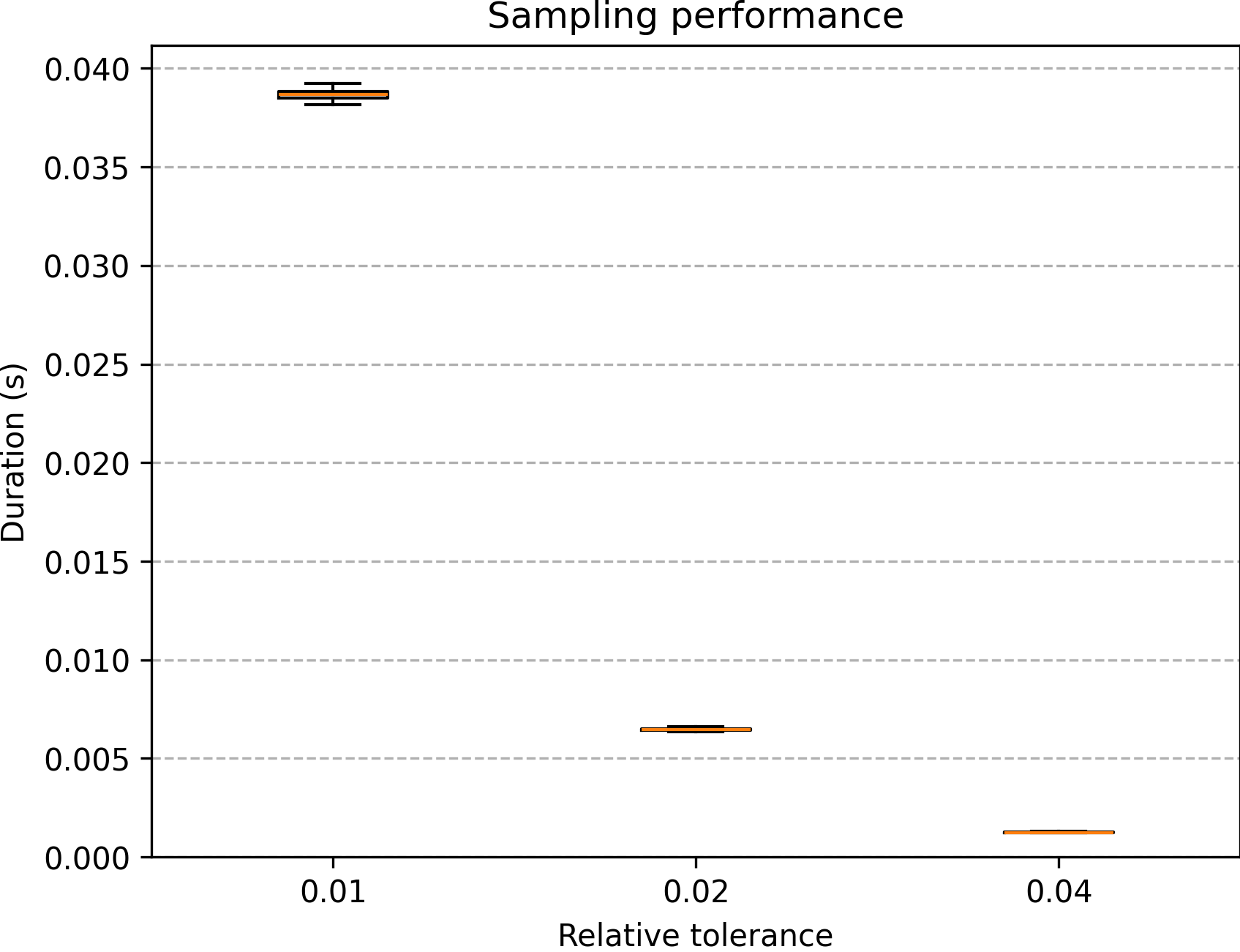}
        \subcaption{Sampling performance. }
    \end{subfigure}
    \\[\baselineskip]
    \begin{subfigure}[t]{0.48\linewidth}
        \includegraphics[width=\linewidth]{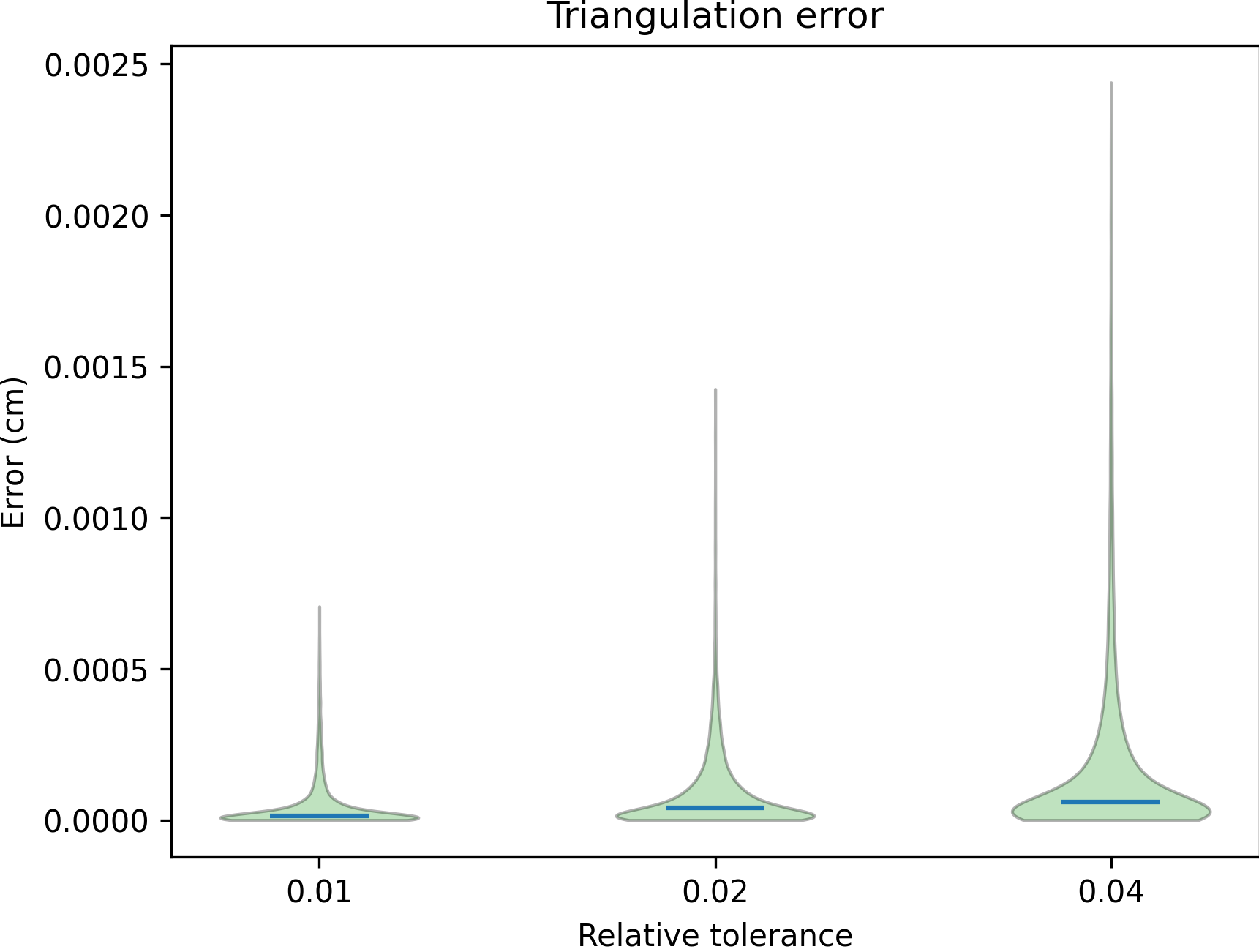}
        \subcaption{Triangulation accuracy. }
    \end{subfigure}
    \hfill
    \begin{subfigure}[t]{0.48\linewidth}
        \includegraphics[width=\linewidth]{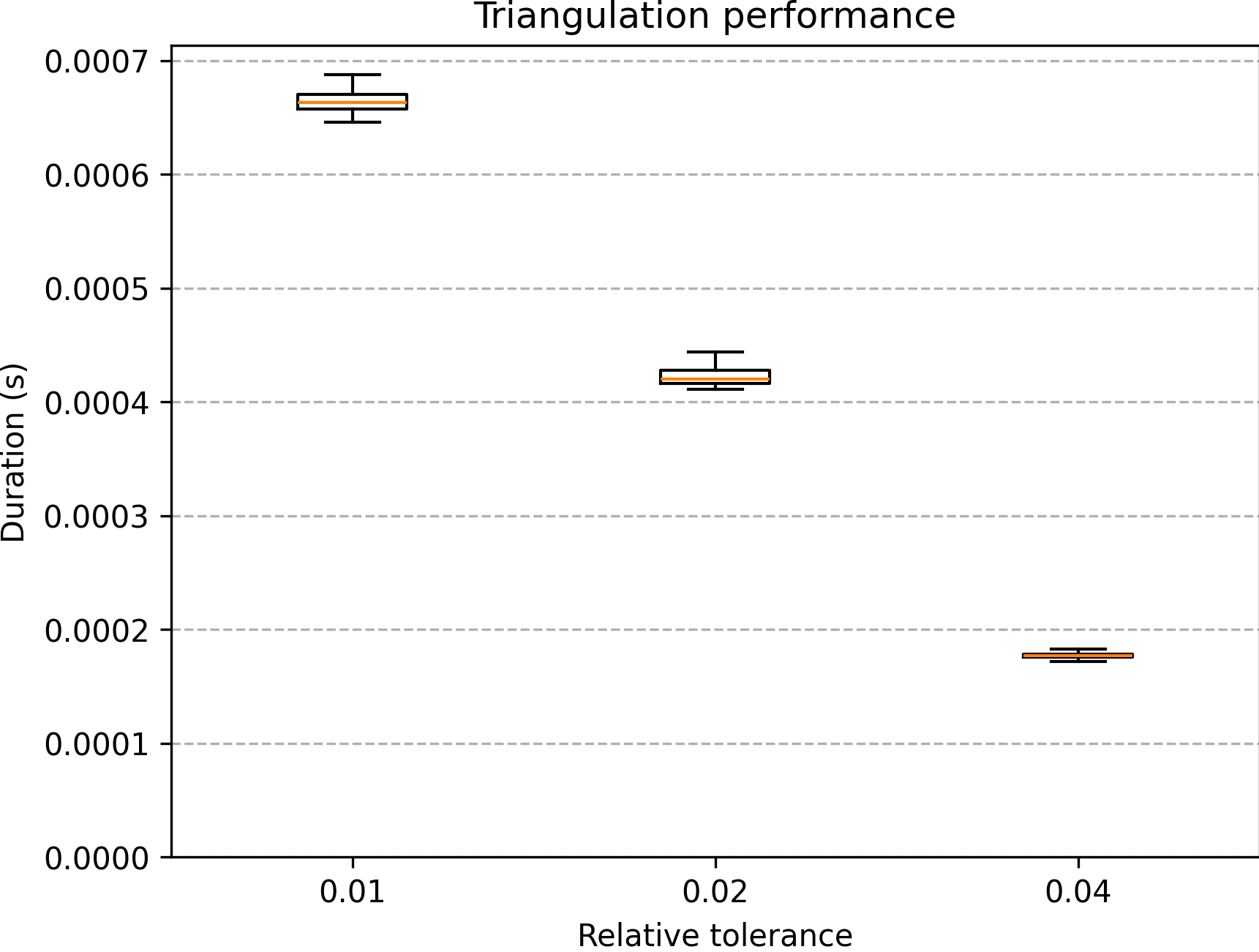}
        \subcaption{Triangulation performance. }
    \end{subfigure}

    \caption{Results for accuracy and performance testing for each distance algorithm.
        Errors are plotted on the left and performance on the right. The minimum distance
    for all results is set to $0.1$ cm. }
    \label{fig:distance_bench}
\end{figure}

Figure~\ref{fig:distance_acc} shows the errors for all three distance functions on a
aortic valve cusp. Here, points on the surface with a ground truth distance of zero
are passed on to the distance algorithms. The same data is also shown as violin plots
in the left column in Figure~\ref{fig:distance_bench}. The triangulation and sampling
are tested with error tolerance settings that aim to achieve results below that
tolerance. For example, a $2\%$ tolerance and minimum distance of $0.1$ cm would give
a maximum error of $0.002$ cm. The minimum distance is set to $0.1$ cm for all
results. The tolerance for minimization is the threshold of relative difference
between the distances of two iterations, where the iteration is terminated when the
distance is below that threshold. Because the tolerance is defined differently for
minimization, other tolerance parameters are used for the comparison.

Of the three algorithms, sampling with a low sampling rate has the highest overall
error rate with a pattern that shows bigger distances between samples. Although the
triangulation has a relatively low error rate overall, there are some visible areas
where the crease of the triangles are visible due to high error rates. The
minimization algorithm has the lowest error rate, with no visible errors. However, the
violin plot does show that there are some outliers. Also, the set tolerance seems to
have little impact on the error, because the errors achieved with $0.001$ and $0.01$
are the same.

Figure~\ref{fig:distance_bench} shows the performance and accuracy of the distance
algorithms when computing the distance of a point to a surface. Generally, the
triangulation algorithm is the fastest, followed by sampling and minimization. As
expected, the sampling and triangulation algorithms' computing time do not vary much
for different probing points, in contrast to minimization. This is likely due to the
fact that sampling and triangulation always perform the same number of calculations
for any point, but minimization can have early termination once a local minimum has been found.

In terms of scaling with tolerance parameters, minimization does not show a big change
in duration when increasing the tolerance. This can be explained by the non-controlled
behavior of the optimization iterations. Sampling, on the other hand, seems to scale
quadratically with the tolerance level, where a tolerance of $0.02$ is roughly $5$
times faster than a tolerance of $0.01$. Triangulation seems to scale linearly with
the tolerance setting.

\subsection{Mesh traversal} \label{sec:mesh_traversal}
To use the surface in a volumetric mesh, all nodes in the mesh that lie in the surface
have to be found. For this we use a mesh traversal algorithm that finds all surface
nodes given one of the distance functions discussed in Section~\ref{sub:distance}. We
have implemented two mesh traversal algorithms: Exhaustive node iteration
(Section~\ref{sub:exhaustive}) and recursive neighbor search (Section~\ref{sub:neighbor}).

\subsubsection{Exhaustive node iteration}\label{sub:exhaustive}
For the exhaustive node iteration the distance $d_i$ of every node $M_i$ to the
surface $S$ is computed and added to the list of surface nodes $M_S$ if it is below
the surface thickness $d_{min}$. The node iteration algorithm is shown in
Algorithm~\ref{alg:node_iteration}. This implementation can easily be parallelized by
dividing the mesh nodes across multiple threads.

\begin{algorithm}[H]
    \caption{Exhaustive node iteration algorithm.}
    \label{alg:node_iteration}

    \begin{algorithmic}
        \Require{Mesh $M$, surface $S$}
        \Ensure{Surface nodes $M_S$}
        \Function{Node iteration}{$M, S$}
        \State $M_S \gets \{\}$
        \ForAll{$M_i \in M$}
        \State $d_i \gets D(M_i, S)$
        \If{$d_i \leq d_{min}$ }
        \State $M_S \gets M_S \cup \{i\}$
        \EndIf
        \EndFor
        \State \Return $M_S$
        \EndFunction
    \end{algorithmic}
\end{algorithm}

\subsubsection{Recursive neighbor search}\label{sub:neighbor}
The recursive neighbor search traversal algorithm makes use of the property that the
surface is one connected piece, i.e., there exists a path on the surface that connects
any point on $S$ to any other point on $S$. This gives that all mesh nodes $M_S$ are
also connected if the mesh density is small enough. The algorithm starts by finding
the first node $M_i$ that is part of the surface by calculating the distance of one
arbitrary point on the surface $S(u_{\start},v_{\start})$ to every node $M_i$ and
taking the node with the lowest distance:
\begin{align}
    i_{\start} = &\argmin_i \|S(u_{\start},v_{\start}) - M_i \| .
\end{align}
This requires that $ \| S(u_{\start},v_{\start}) - M_{i_{\start}}  \| \leq d_{\min}$
to ensure $M_{i_{\start}}$ can be marked as a surface node.

After the first point is found, a recursive neighbor search is used to find all other
mesh nodes that belong to the surface. The recursive neighbor search traversal
algorithm is shown in Algorithm~\ref{alg:neighbor_traversal}. This algorithm can be
parallelized by having multiple threads process items from the queue until it is empty.

\begin{algorithm}[H]
    \caption{Recursive neighbor search algorithm.}
    \label{alg:neighbor_traversal}

    \begin{algorithmic}
        \Require{Mesh $M$, surface $S$}
        \Ensure{Surface nodes $M_S$}
        \Function{Recursive neighbor search}{$M, S$}
        \State $i_{\start} = \argmin\limits_i  \|S(u_{start},v_{start}) - M_i \| $
        \State \textbf{Assert} $\| S(u_{\start},v_{\start}) - M_{i_{\start}}  \| \leq t_S$
        \State $Q \gets $ Queue()
        \State $Q.$put$(M_{i_{\start}})$
        \State $O \gets \{ i_{\start} \}$
        \State $M_S \gets \{ i_{\start} \}$
        \While{$Q$ is not empty}
        \State $i \gets Q.$get$()$
        \If{$D(M_i, S) < t_S$}
        \State $M_S \gets M_S \cup \{ M_i \}$
        \ForAll{Neighbors $M_j$ of $M_i$}
        \If{$j$ not in $O$}
        \State $O \gets O \cup \{ j \}$
        \State $Q.\text{put}(j)$
        \EndIf
        \EndFor
        \EndIf
        \EndWhile
        \State \Return $M_S$
        \EndFunction
    \end{algorithmic}
\end{algorithm}

\subsubsection{Performance evaluation}

\begin{figure}[htbp]
    \centering
    \begin{subfigure}[t]{0.48\linewidth}
        \includegraphics[width=\linewidth]{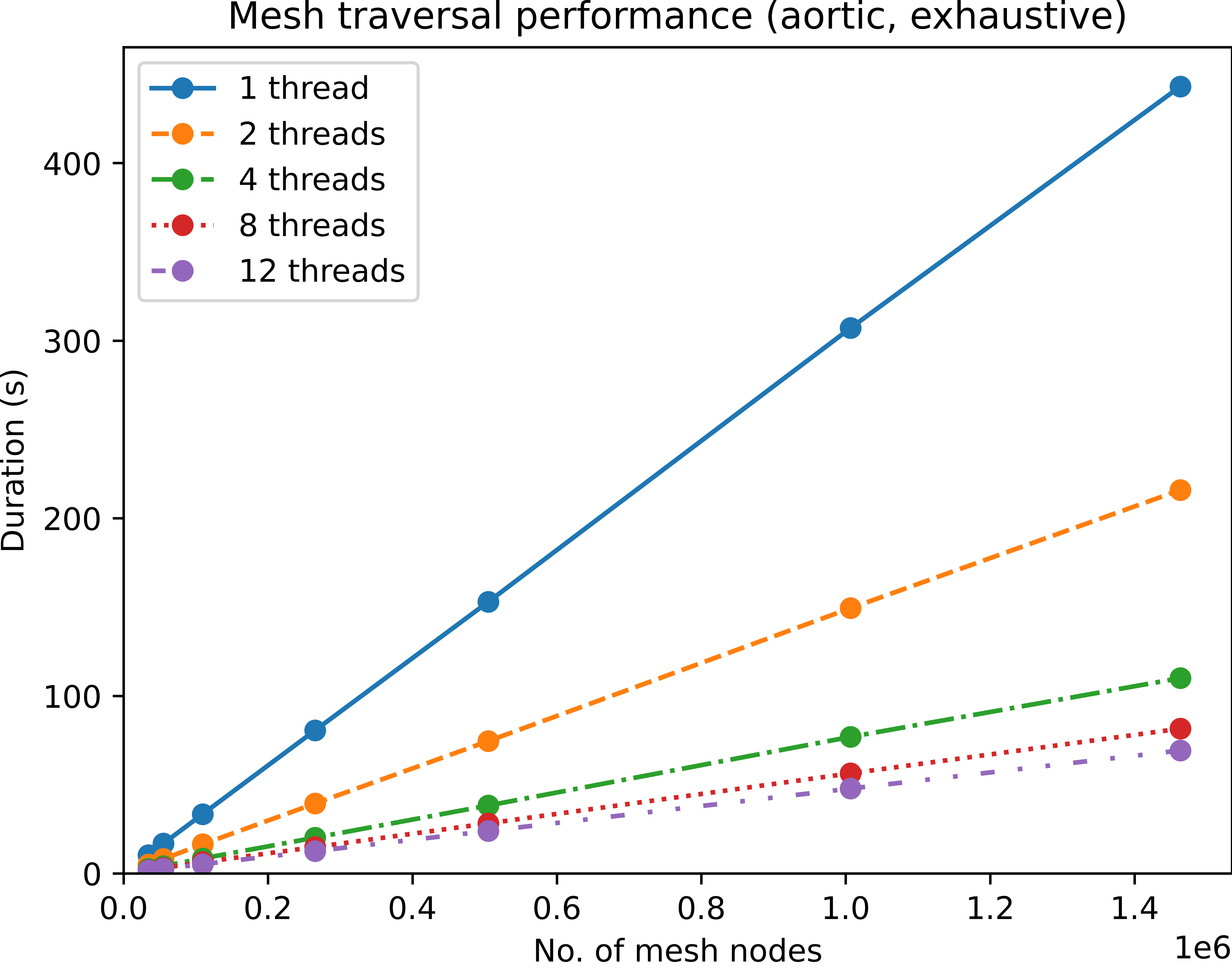}
        \caption{Exhaustive search with aortic valve. }
    \end{subfigure}
    \begin{subfigure}[t]{0.48\linewidth}
        \includegraphics[width=\linewidth]{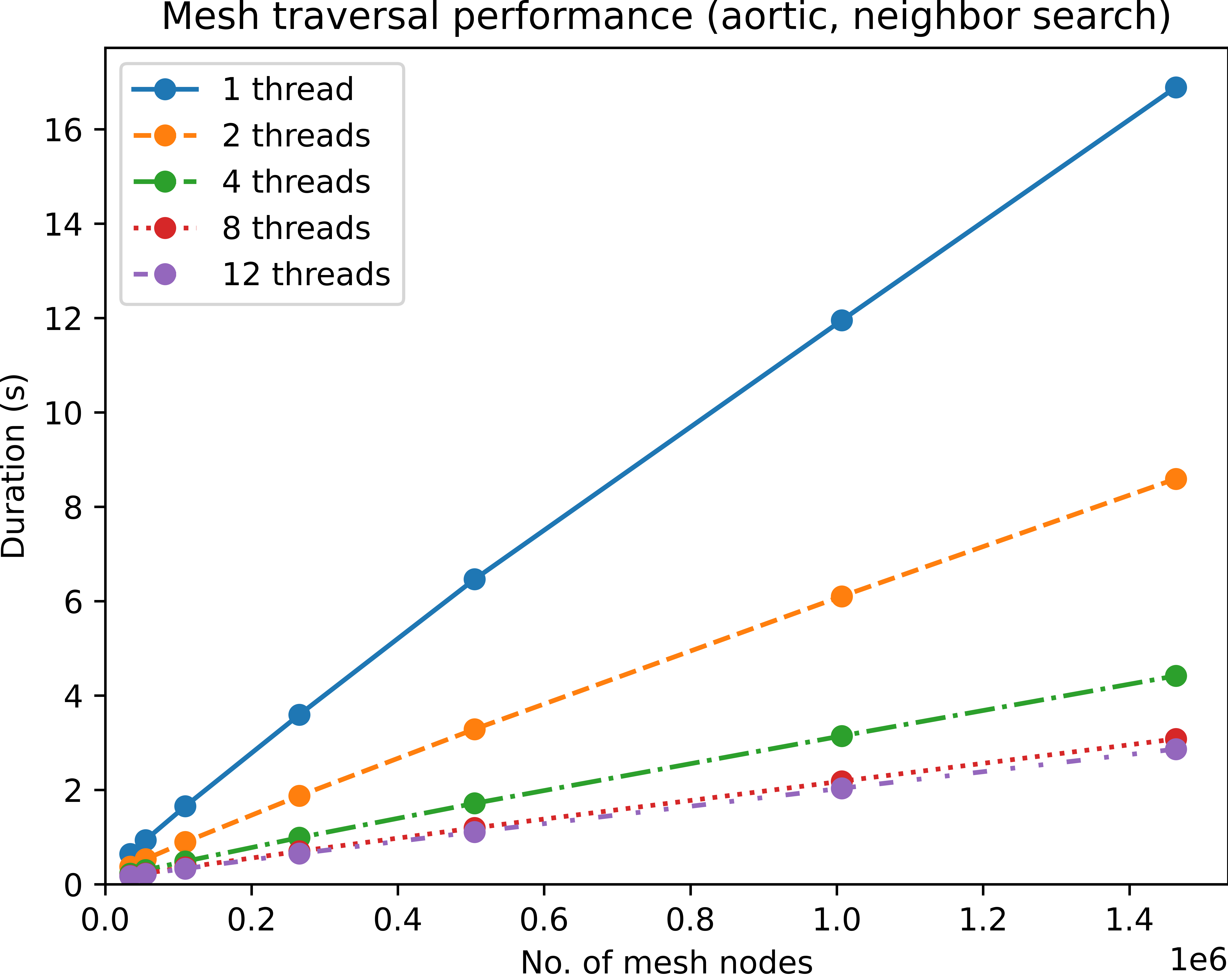}
        \caption{Recursive neighbor search with aortic valve. }
    \end{subfigure}
    \\[\baselineskip]
    \begin{subfigure}[b]{0.48\linewidth}
        \includegraphics[width=\linewidth]{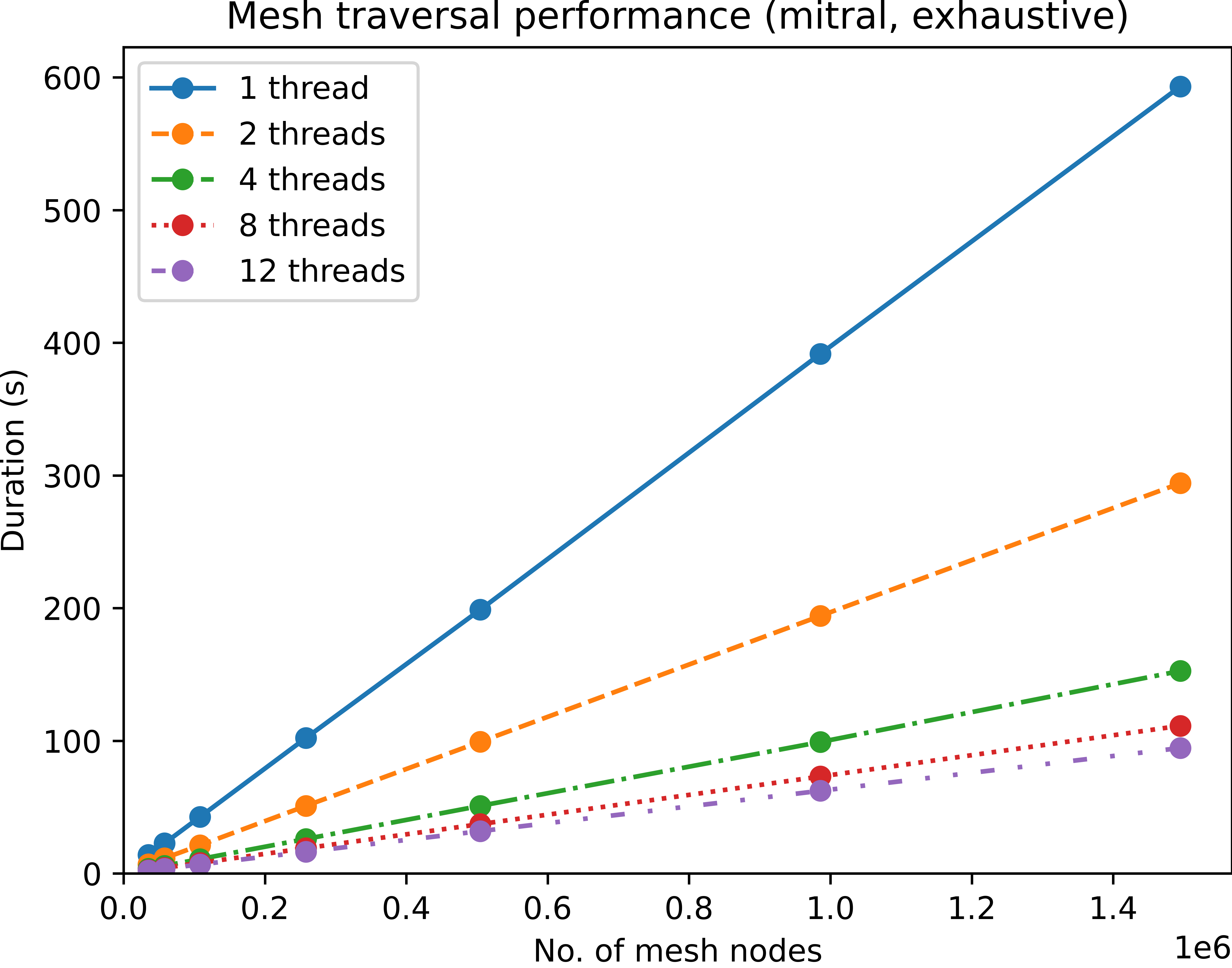}
        \caption{Exhaustive search with mitral valve. }
    \end{subfigure}
    \begin{subfigure}[b]{0.48\linewidth}
        \includegraphics[width=\linewidth]{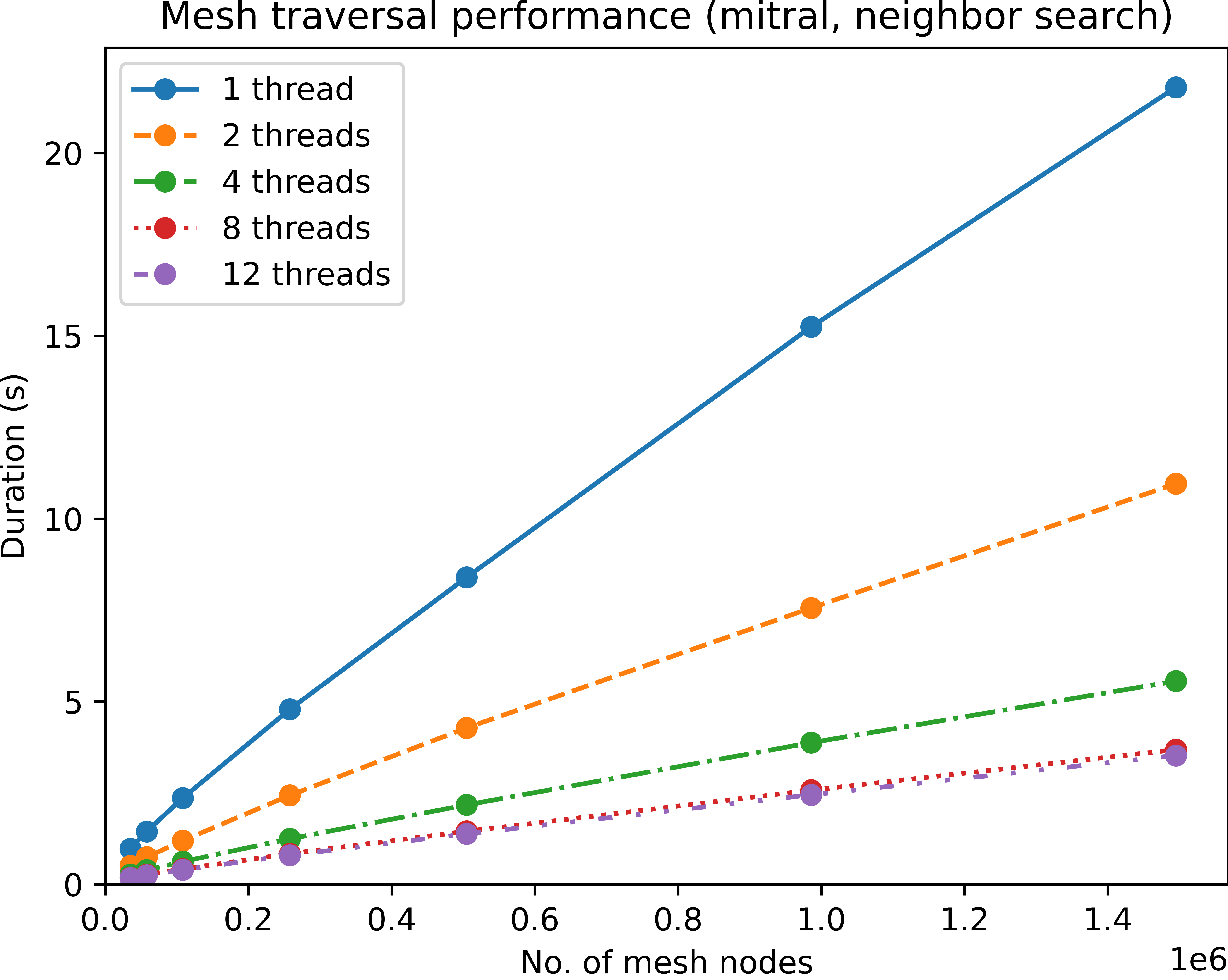}
        \caption{Recursive neighbor search with mitral valve. }
    \end{subfigure}
    \caption{Performance results of sequential mesh traversal functions tested with
    triangulation on the aortic valve and mitral valve.}
    \label{fig:traverse_perf}
\end{figure}

Figure~\ref{fig:traverse_perf} shows the performance of the mesh traversal functions
on the aortic valve and mitral valve using the triangulation method with 5\%
tolerance. Simple cylinder meshes are used for testing, where only the mesh density
changes with the number of nodes and the physical size stays the same. The aortic
valve and mitral valve are tested with meshes that fit the respective valve sizes.
Both sequential and parallel implementations are tested. The data is also shown in
Table~\ref{tab:traverse_scaling_aortic} and Table~\ref{tab:traverse_scaling_mitral}.

Both mesh traversal methods scale linearly with the number of mesh nodes. The
recursive neighbor search algorithm is $18$ to $24$ times faster than the exhaustive
algorithm when comparing the sequential implementations. The scaling with multiple
threads is approximately equal for both algorithms, where nearly linear speedup is
achieved with up to $4$, but the relative speedup drops with higher numbers of
threads. For $12$ threads a $6\times$ speedup is achieved. When comparing the speeds
of the aortic valve cusp with the mitral valve, we can see that the mitral valve is
about $30\%$ slower than the aortic valve on average. This can be explained by the
fact that the mitral valve is larger than the aortic valve, which gives that the
mitral valve surface consists of more points/quads that need to be checked for
distance calculations.

\begin{table}[H]
    \centering
    \begin{tabular}{|ll|ll|ll|}
\hline
\multirow{2}{*}{\textbf{No. nodes}} & \multirow{2}{*}{\textbf{No. threads}} 
& \multicolumn{2}{c|}{\textbf{Exhaustive search}}& \multicolumn{2}{c|}{\textbf{Recursive neighbor search}}\\ 
& & \textbf{Duration (s)} & \textbf{Speedup}& \textbf{Duration (s)} & \textbf{Speedup}\\
\hline 
\multirow{5}{*}{109464}& 1 & 33.420 & -& 1.655 & -\\ 
& 2 & 16.461 & 2.030& 0.896 & 1.847\\ 
& 4 & 8.475 & 3.943& 0.487 & 3.401\\ 
& 8 & 6.395 & 5.226& 0.359 & 4.610\\ 
& 12 & 5.213 & 6.411& 0.333 & 4.962\\ 
\hline 
\multirow{5}{*}{265277}& 1 & 80.608 & -& 3.593 & -\\ 
& 2 & 39.413 & 2.045& 1.874 & 1.917\\ 
& 4 & 20.192 & 3.992& 0.990 & 3.630\\ 
& 8 & 14.880 & 5.417& 0.699 & 5.137\\ 
& 12 & 12.609 & 6.393& 0.652 & 5.508\\ 
\hline 
\multirow{5}{*}{504929}& 1 & 152.917 & -& 6.464 & -\\ 
& 2 & 74.513 & 2.052& 3.288 & 1.966\\ 
& 4 & 38.272 & 3.996& 1.716 & 3.768\\ 
& 8 & 28.230 & 5.417& 1.196 & 5.406\\ 
& 12 & 23.932 & 6.389& 1.110 & 5.823\\ 
\hline 
\multirow{5}{*}{1006781}& 1 & 307.133 & -& 11.950 & -\\ 
& 2 & 149.403 & 2.056& 6.103 & 1.958\\ 
& 4 & 76.925 & 3.993& 3.144 & 3.800\\ 
& 8 & 56.388 & 5.447& 2.184 & 5.472\\ 
& 12 & 47.824 & 6.422& 2.035 & 5.872\\ 
\hline 
\multirow{5}{*}{1463509}& 1 & 443.155 & -& 16.889 & -\\ 
& 2 & 215.921 & 2.052& 8.592 & 1.966\\ 
& 4 & 110.017 & 4.028& 4.417 & 3.823\\ 
& 8 & 81.550 & 5.434& 3.081 & 5.482\\ 
& 12 & 69.277 & 6.397& 2.865 & 5.894\\ 
\hline
\end{tabular}
    \caption{Scaling of mesh traversal functions for aortic valve with multithreading. }
    \label{tab:traverse_scaling_aortic}
\end{table}

\begin{table}[H]
    \centering
    \begin{tabular}{|ll|ll|ll|}
\hline
\multirow{2}{*}{\textbf{No. nodes}} & \multirow{2}{*}{\textbf{No. threads}} 
& \multicolumn{2}{c|}{\textbf{Exhaustive search}}& \multicolumn{2}{c|}{\textbf{Recursive neighbor search}}\\ 
& & \textbf{Duration (s)} & \textbf{Speedup}& \textbf{Duration (s)} & \textbf{Speedup}\\
\hline 
\multirow{5}{*}{108233}& 1 & 42.718 & -& 2.359 & -\\ 
& 2 & 21.306 & 2.005& 1.193 & 1.978\\ 
& 4 & 10.889 & 3.923& 0.620 & 3.805\\ 
& 8 & 8.007 & 5.335& 0.416 & 5.674\\ 
& 12 & 6.813 & 6.270& 0.395 & 5.974\\ 
\hline 
\multirow{5}{*}{257845}& 1 & 102.086 & -& 4.787 & -\\ 
& 2 & 50.927 & 2.005& 2.433 & 1.967\\ 
& 4 & 26.008 & 3.925& 1.246 & 3.841\\ 
& 8 & 19.067 & 5.354& 0.837 & 5.718\\ 
& 12 & 16.307 & 6.260& 0.793 & 6.037\\ 
\hline 
\multirow{5}{*}{504683}& 1 & 198.841 & -& 8.393 & -\\ 
& 2 & 99.259 & 2.003& 4.278 & 1.962\\ 
& 4 & 50.964 & 3.902& 2.172 & 3.865\\ 
& 8 & 37.224 & 5.342& 1.450 & 5.788\\ 
& 12 & 31.897 & 6.234& 1.380 & 6.084\\ 
\hline 
\multirow{5}{*}{985877}& 1 & 391.664 & -& 15.246 & -\\ 
& 2 & 194.165 & 2.017& 7.557 & 2.017\\ 
& 4 & 99.121 & 3.951& 3.873 & 3.936\\ 
& 8 & 73.097 & 5.358& 2.574 & 5.924\\ 
& 12 & 62.320 & 6.285& 2.448 & 6.228\\ 
\hline 
\multirow{5}{*}{1495173}& 1 & 593.225 & -& 21.797 & -\\ 
& 2 & 294.156 & 2.017& 10.955 & 1.990\\ 
& 4 & 152.668 & 3.886& 5.558 & 3.922\\ 
& 8 & 111.331 & 5.328& 3.683 & 5.919\\ 
& 12 & 94.585 & 6.272& 3.525 & 6.184\\ 
\hline
\end{tabular}
    \caption{Scaling of mesh traversal functions for mitral valve with multithreading.}
    \label{tab:traverse_scaling_mitral}
\end{table}

\begin{figure}[htbp]
    \centering
    \begin{subfigure}{0.48\linewidth}
        \includegraphics[width=\linewidth]{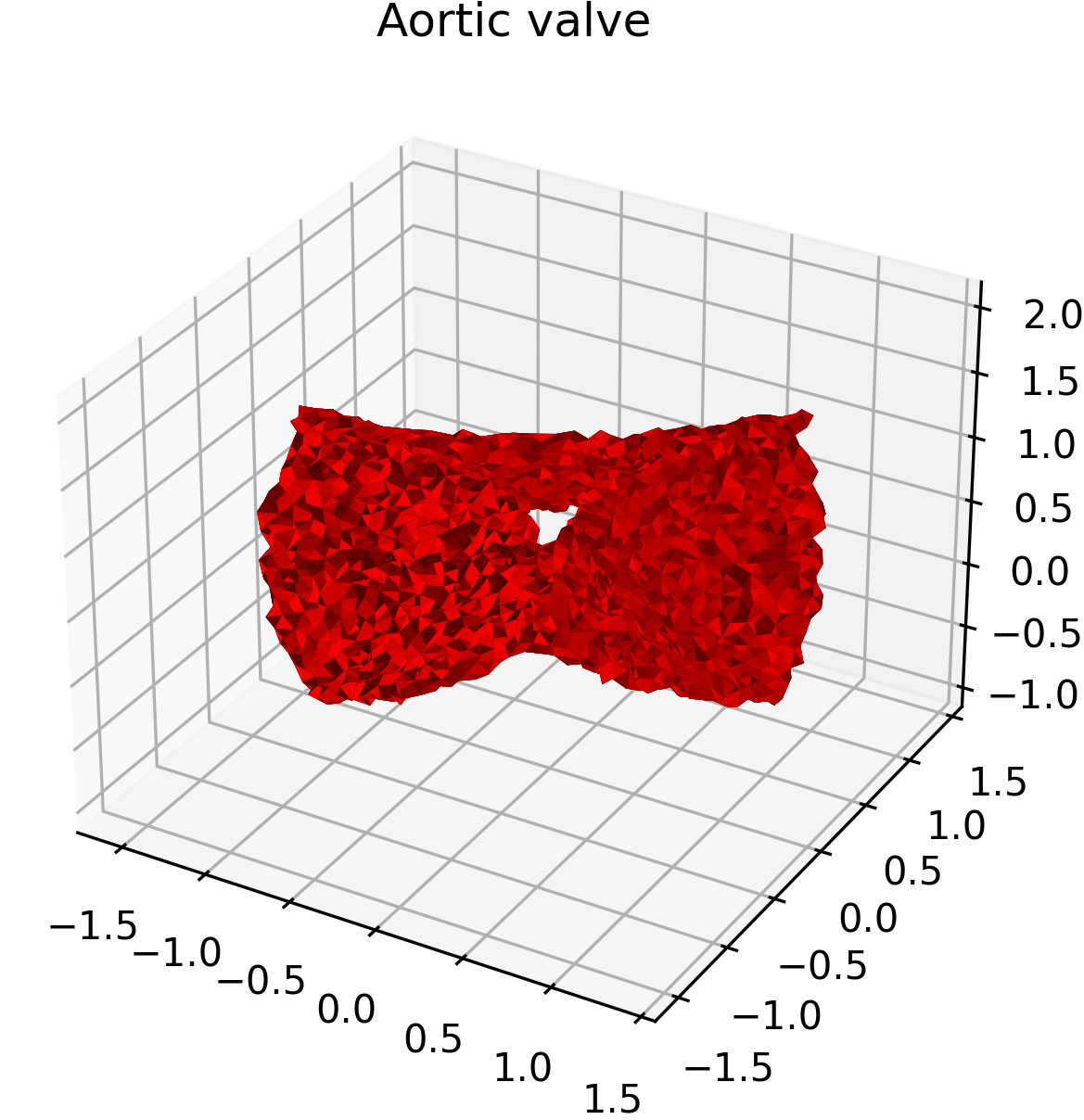}
        \caption{Aortic valve side view. }
    \end{subfigure}
    \begin{subfigure}{0.48\linewidth}
        \includegraphics[width=\linewidth]{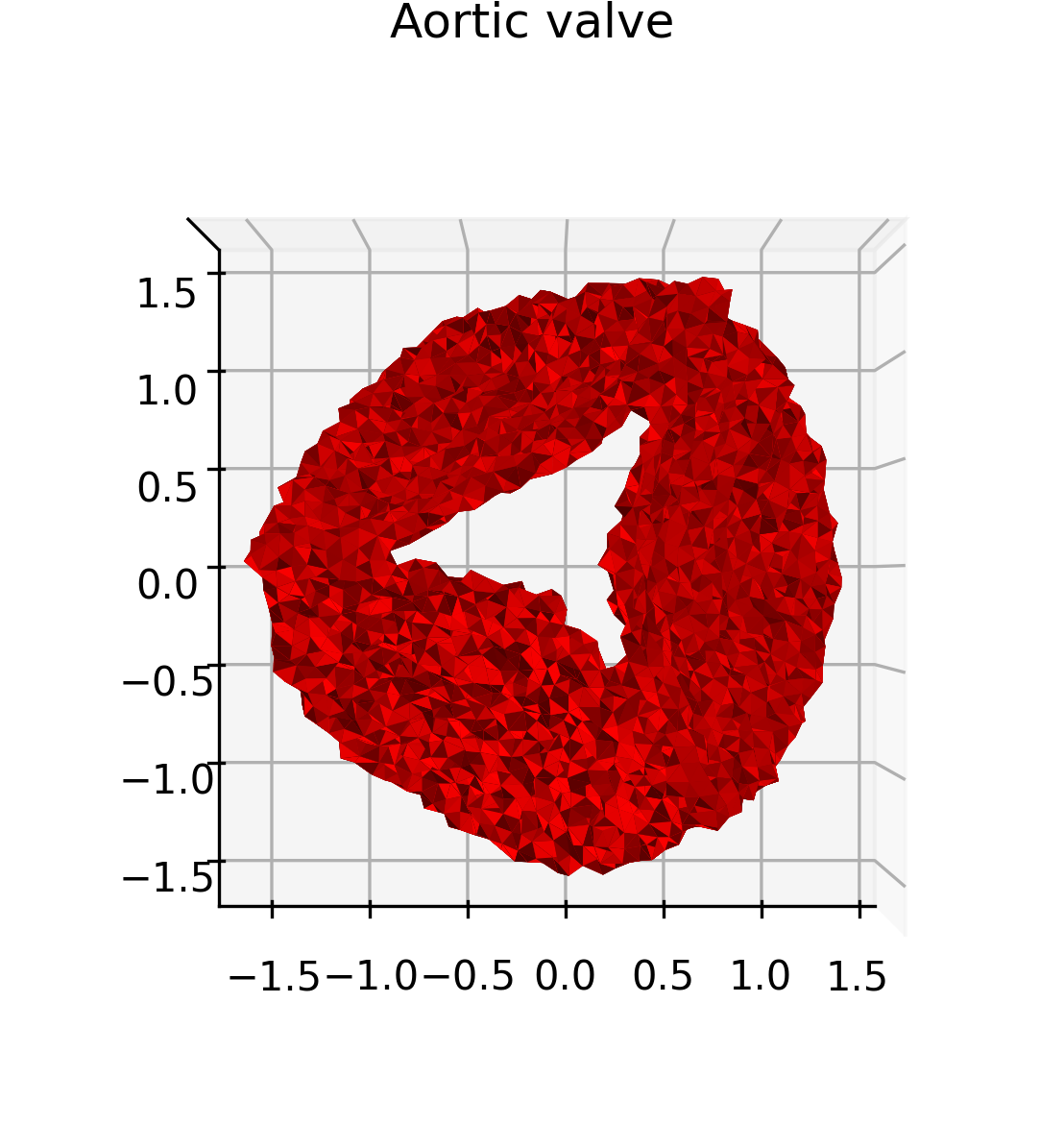}
        \caption{Aortic valve top view. }
    \end{subfigure}
    \\[\baselineskip]
    \begin{subfigure}{0.48\linewidth}
        \includegraphics[width=\linewidth]{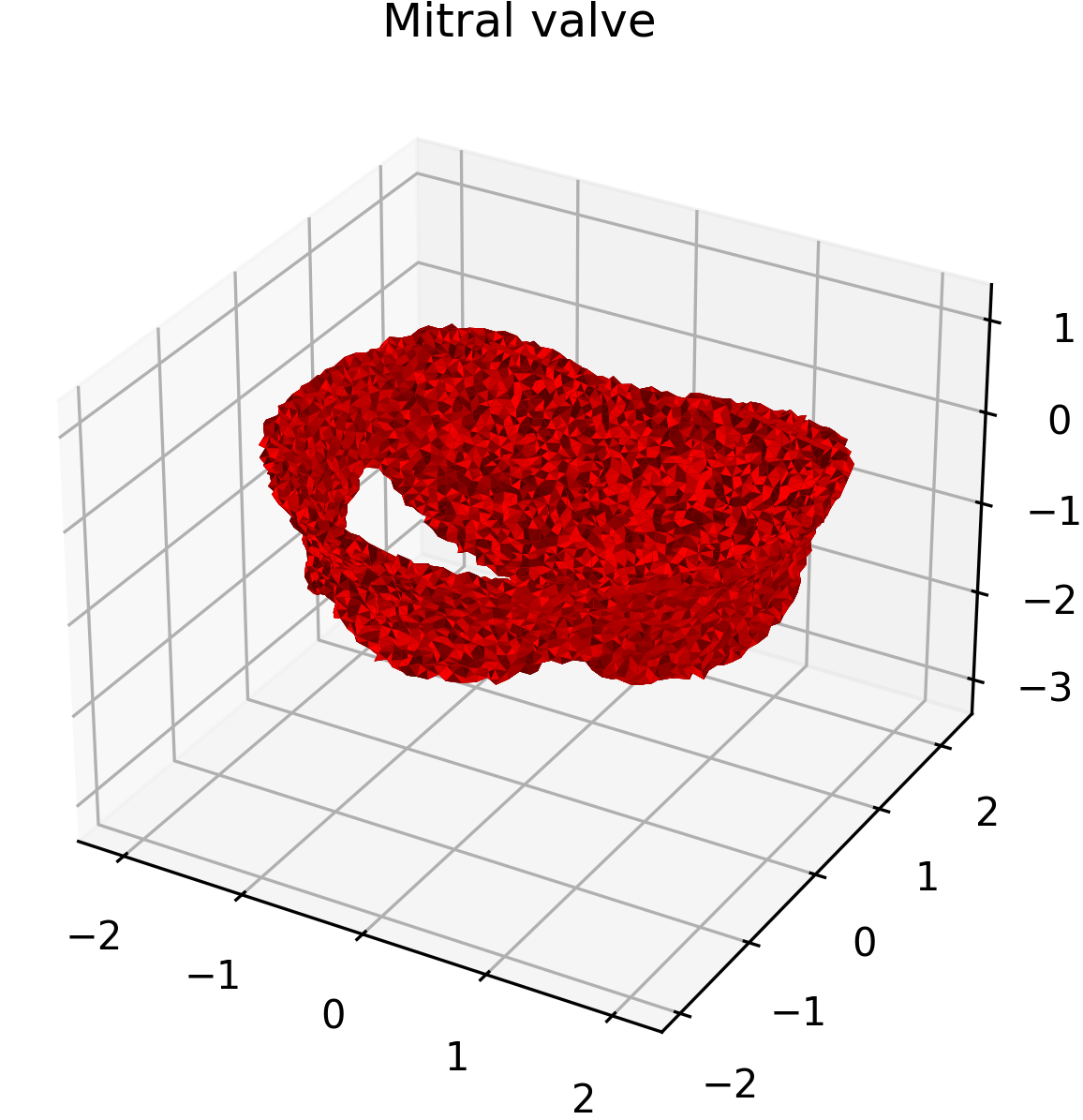}
        \caption{Mitral valve side view. }
    \end{subfigure}
    \begin{subfigure}{0.48\linewidth}
        \includegraphics[width=\linewidth]{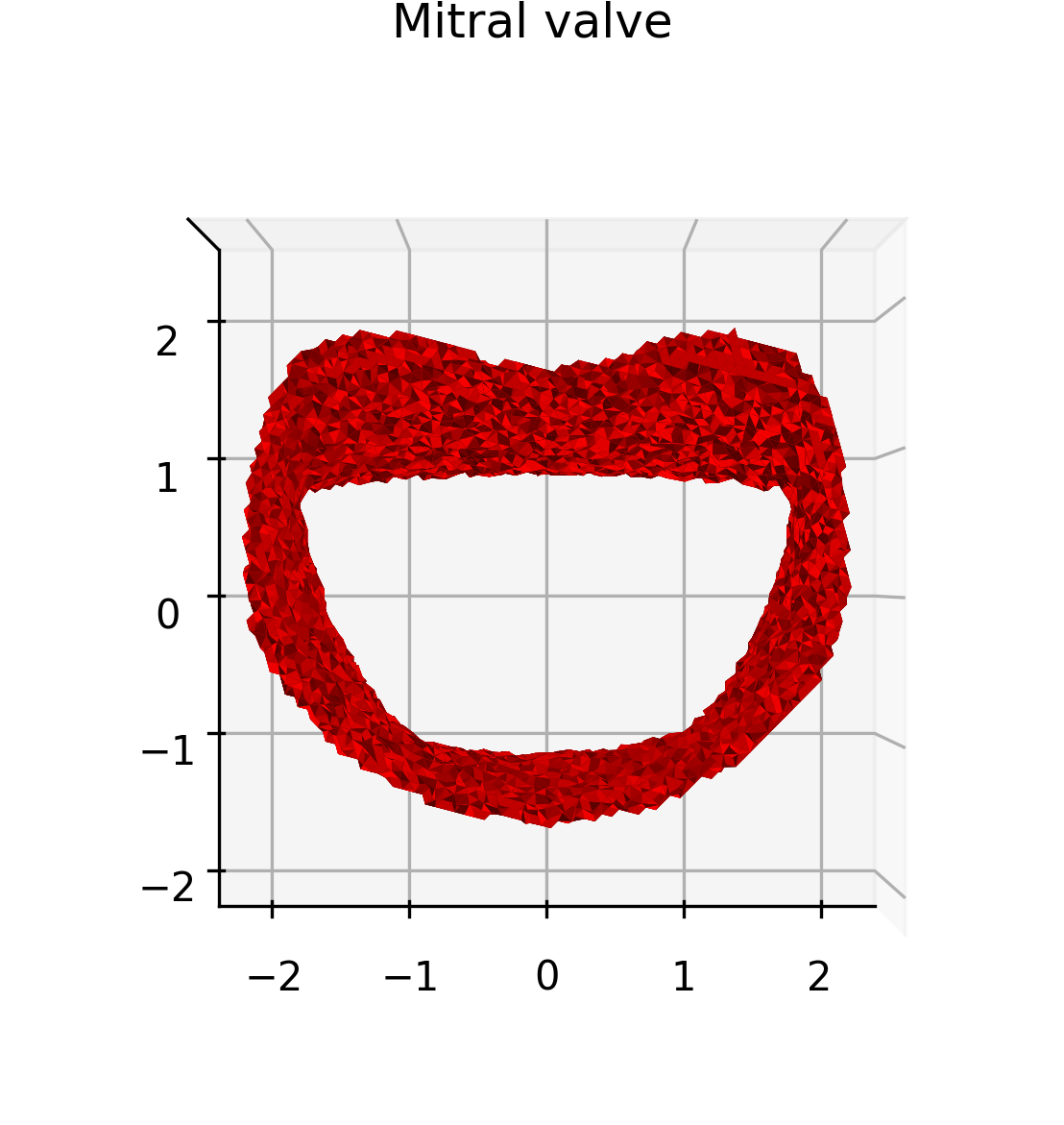}
        \caption{Mitral valve top view. }
    \end{subfigure}
    \caption{Results of mesh traversal algorithms with aortic and mitral valves.}
    \label{fig:final_results}

\end{figure}

Figure~\ref{fig:final_results} shows the tetrahedra of a volumetric mesh for the
aortic and mitral valves. These are found by recursive neighbor search using the
triangulation distance function.

\section{Conclusions} \label{sec:conclusions}

In this work, we presented a fast and flexible pipeline for generating 3D flow
obstacles from parametric surface models of the aortic and mitral valves. The proposed
approach uses adaptive surface representations, efficient distance algorithms, and a
recursive mesh traversal strategy to significantly reduce computational cost compared
to an exhaustive search. Among the tested distance algorithms, triangulation offers
the best trade-off between speed and accuracy. The recursive neighbor search further
accelerates the identification of mesh nodes belonging to the valve surface,
especially for large meshes.

The method is demonstrated on both an aortic valve model and a new mitral valve,
showing its generality and efficiency. While direct comparisons with other methods are
limited due to the lack of similar published pipelines, the presented results indicate
that the approach is suitable for rapid valve shape updates in inverse estimation and
simulation workflows. Future work may focus on further optimizing the distance
algorithms, and integrating the heart valves into a heart model.

\section*{Acknowledgements} \label{sec:ack}

C.B. acknowledges the funding from the European Research Council (ERC) under the
European Union’s Horizon 2020 research and innovation program (grant agreement No
852544 - CardioZoom).

\section*{Ethics Statement}

None.

\appendix

\section{Parametric modeling of the mitral valve} \label{app:mitralvalve}

\subsection{Anatomy}

\subsubsection{Annulus}
\begin{figure}[htbp]
    \centering
    \includegraphics[width=0.6\linewidth]{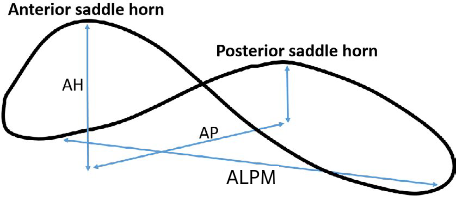}
    \caption{Saddle shape of the annulus. Image taken from \cite{oliveira_geometric_2020}. }
    \label{fig:annulus_saddle}
\end{figure}

The shape of the annulus in the closed position can be described as a hyperbolic
paraboloid, which is similar to the shape of a horse saddle or mitre (hence the name).
The peaks or horns are at the anterior and posterior tips of the annulus, as is shown
in Figure~\ref{fig:annulus_saddle}. The height of the anterior horn is defined by the
anterior height (AH) and the width and depth of the annulus are defined by the
antero-posterior (AP) and the anterolateral-posteromedial (ALPM) distances. There is
an indent in the anterior near the aortic valve.

\begin{figure}[htbp]
    \centering
    \includegraphics[width=0.6\linewidth]{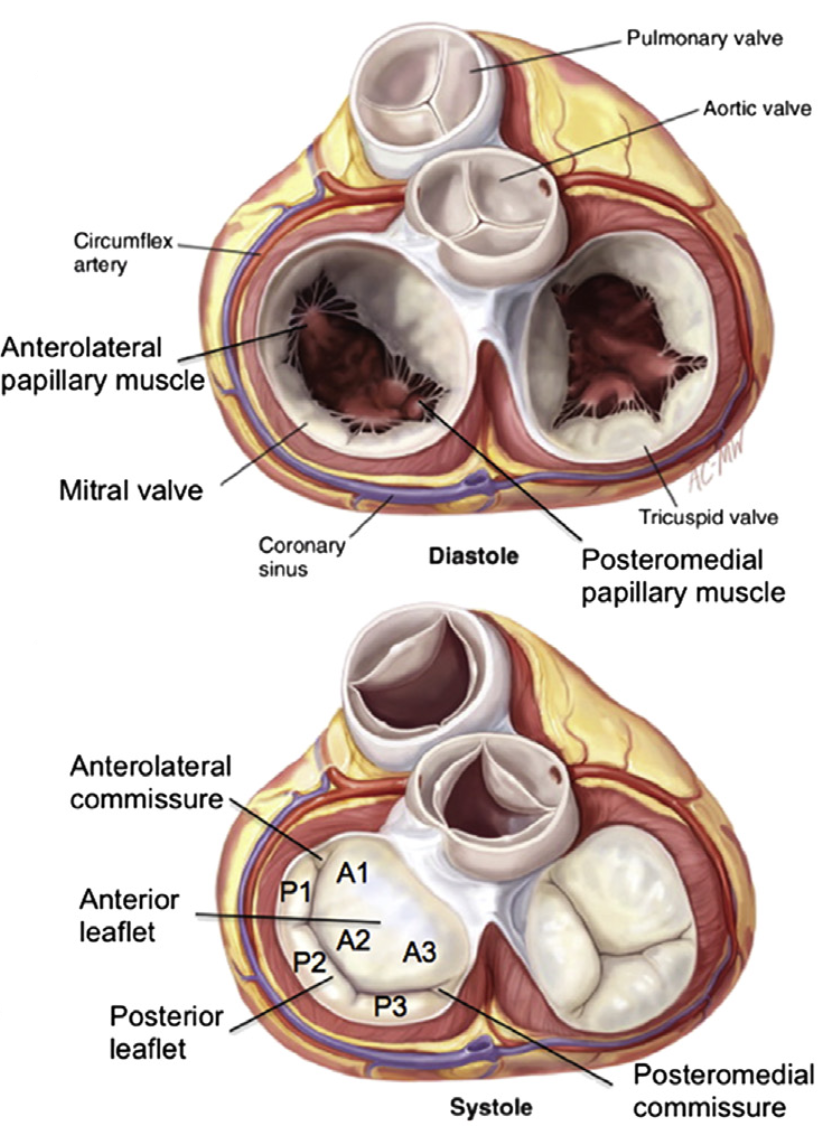}
    \caption{Top view of the heart in diastole and systole with the location and
    features of the mitral valve. Image taken from \cite{dal-bianco_anatomy_2013}.}
    \label{fig:top_heart}
\end{figure}

When the valve opens, the annulus' shape changes: The annulus flattens and becomes
planar, but also more circular and the posterior part dilates. The anterior part of
the annulus does not dilate because there is thicker fibrous tissue between the mitral
and aortic valves \cite{mccarthy_anatomy_2010}. This tissue is pictured in white in
Figure~\ref{fig:top_heart}.

\subsubsection{Leaflets}\label{sec:mitral_leaflets}
The anterior leaflet is the largest leaflet of the two and covers most of the surface
area of the mitral valve when closed, as can be seen in Figure~\ref{fig:top_heart}. It
has a single large scallop. The posterior leaflet is smaller and has three scallops.
The areas attached to the annulus between the anterior and posterior leaflets are the
anterolateral and posteromedial commissures.


\subsection{Previously reported mitral valve geometric models}

First, in \cite{shen_geometric_2017} a parametric model of the mitral valve in the
open state was  proposed, where the annulus consists of two half ellipses that model
the anterior and posterior parts. The anterior leaflet and the three scallops on the
posterior leaflet are each defined as half ellipses, where their widths and lengths
can be set individually. 

In \cite{thibaut_use_2019} a simple parametric model of the mitral valve was
introduced. In the model a planar D-shape is used for the annulus and the ratio
between AP and ALPM is fixed. The leaflets curve is modeled by constructing B-splines
between points on the leaflets. The leaflets are positioned under the annulus
perimeter with a small incline. 

Although the models discussed above capture the general shape of the mitral valve,
they lack the flexibility and complexity to accurately model a patient's mitral valve
due to their simplicity.



\subsection{Model overview}\label{sec:mitral_overview}
In this section we discuss the most important design decisions made when implementing
the mitral valve model. We first discuss some general choices and assumptions, and
then focus on the annulus, leaflets and the surface in each subsection.

The model focuses only on the mitral valve itself with the annulus and the leaflets.
Thus, the left ventricle and atrium are not in the model. The chordae tendineae are
also not included in the mitral valve model. Without loss of generality, the model is
positioned such that the AP is aligned on the y axis and it is assumed that the mitral
valve is symmetrical around the y-axis to reduce the number of parameters.

The mitral valve model uses 18 parameters to define its shape, which are listed in
Table~\ref{tab:parameters} with a brief explanation for each one.

Figures of the model curves in closed and opened state with all control points can be
seen in Figures~\ref{fig:annulus_closed} and~\ref{fig:annulus_open} respectively and
the mitral valve model with surfaces shown in an open state in
Figure~\ref{fig:mitral_valve} and closed in Figure~\ref{fig:surfaces_closed}.

\begin{table*}[htbp]
    \centering
    \begin{tabular}{|p{0.3\textwidth} p{0.05\textwidth} p{0.65\textwidth}|} \hline
        Parameter&  Symbol& Explanation\\ \hline
        \multicolumn{3}{|l|}{\textbf{Annulus}}\\
        anterior\_ratio            & $r_{\mathrm{at}}$ & The ratio between the
        circumferences of the anterior part of the annulus and the whole annulus.\\
        AP\_diameter               & $d_{\mathrm{ap}}$ &The length of the
        antero-posterior or the depth of the annulus\\
        AL-PM\_diameter            & $d_{\mathrm{alpm}}$ &The maximum width of the annlus\\
        ALPM\_position\_ratio      & $r_{\mathrm{alpm}}$ &The position of the maximum
        width relative to AP\_diameter\\
        IT\_distance               & $d_{\mathrm{it}}$ &The distance between the
        trigones, which is equal to the width of the indent\\
        indent\_depth              & $d_{\mathrm{id}}$ &The depth of the indent\\
        anterior\_horn\_height     & $h_{\mathrm{at}}$ &The height of the anterior
        annulus point above the y axis\\
        posterior\_horn\_height    & $h_{\mathrm{po}}$ &The height of the posterior
        annulus point above the y axis\\
        \multicolumn{3}{|l|}{\textbf{Leaflets}}\\
        anterior\_length                   & $l_{\mathrm{at}}$ &Length of the anterior leaflet\\
        posteromedial\_commissure\_height  & $h_{\mathrm{pm}}$ &Length of the commissures\\
        posterior\_scallop\_length1        & $l_{\mathrm{ps1}}$ &Length of the
        posterior side scallops\\
        posterior\_scallop\_length2        & $l_{\mathrm{ps2}}$ &Length of the
        posterior middle scallop\\
        leaflet\_arc\_angle                & $\alpha_{\mathrm{a}}$ &The arc angle of
        the curves used for the leaflets\\
        leaflet\_inward\_angle             & $\alpha_{\mathrm{in}}$ &The angle the
        leaflets curve inwards when open, where 0 is straight down\\
        \multicolumn{3}{|l|}{\textbf{Surface}}\\
        surface\_angle                     & $\alpha_{\mathrm{s}}$ &The angle between
        the surface and a straight line between the annulus and leaflets\\
        triangle\_size\_ratio              & $ r_{\mathrm{tr}}$ &Size of the triangle
        at the commissure as ratio of total annulus circumference\\
        \hline
    \end{tabular}
    \caption{Overview of all parameters used in the mitral valve model. }
    \label{tab:parameters}
\end{table*}

\subsubsection{Annulus}
\begin{figure}[htbp]
    \centering
    \begin{subfigure}{0.49\linewidth}
        \includegraphics[width=\linewidth]{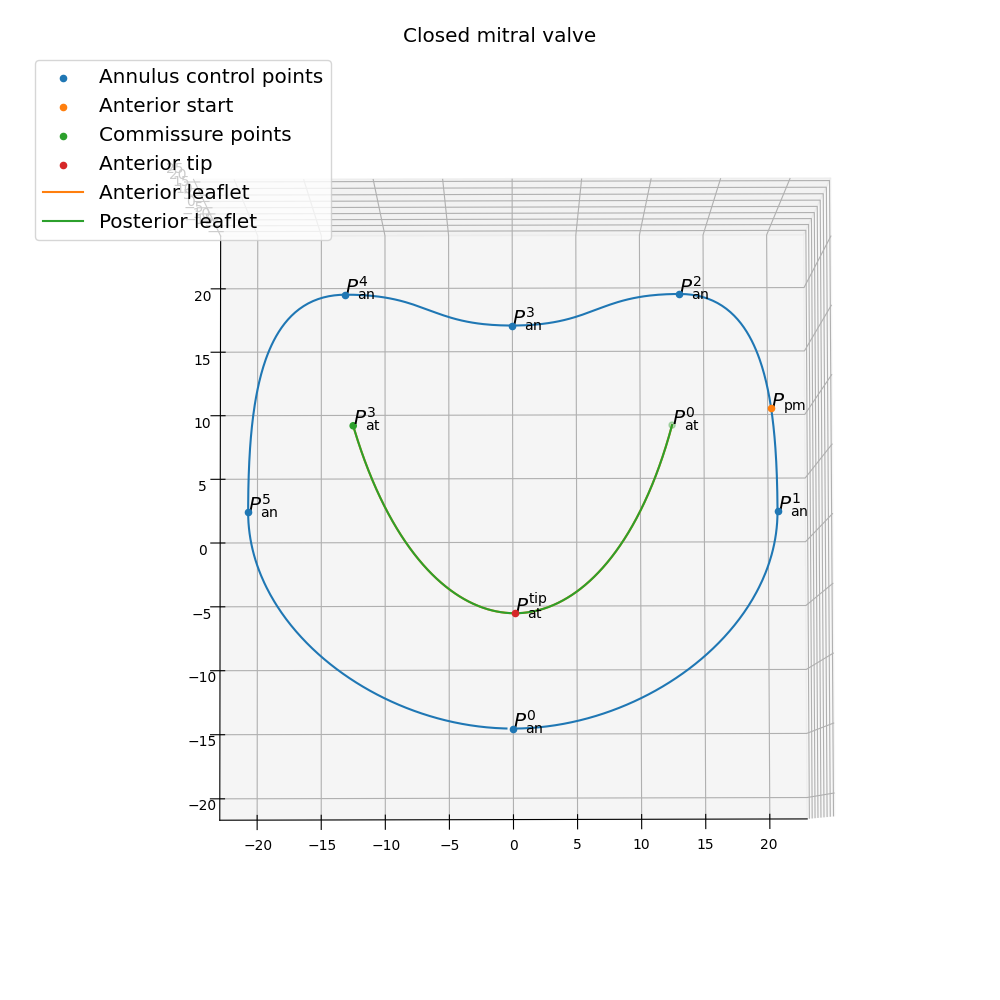}
        \subcaption{Top view}
    \end{subfigure}
    \begin{subfigure}{0.49\linewidth}
        \includegraphics[width=\linewidth]{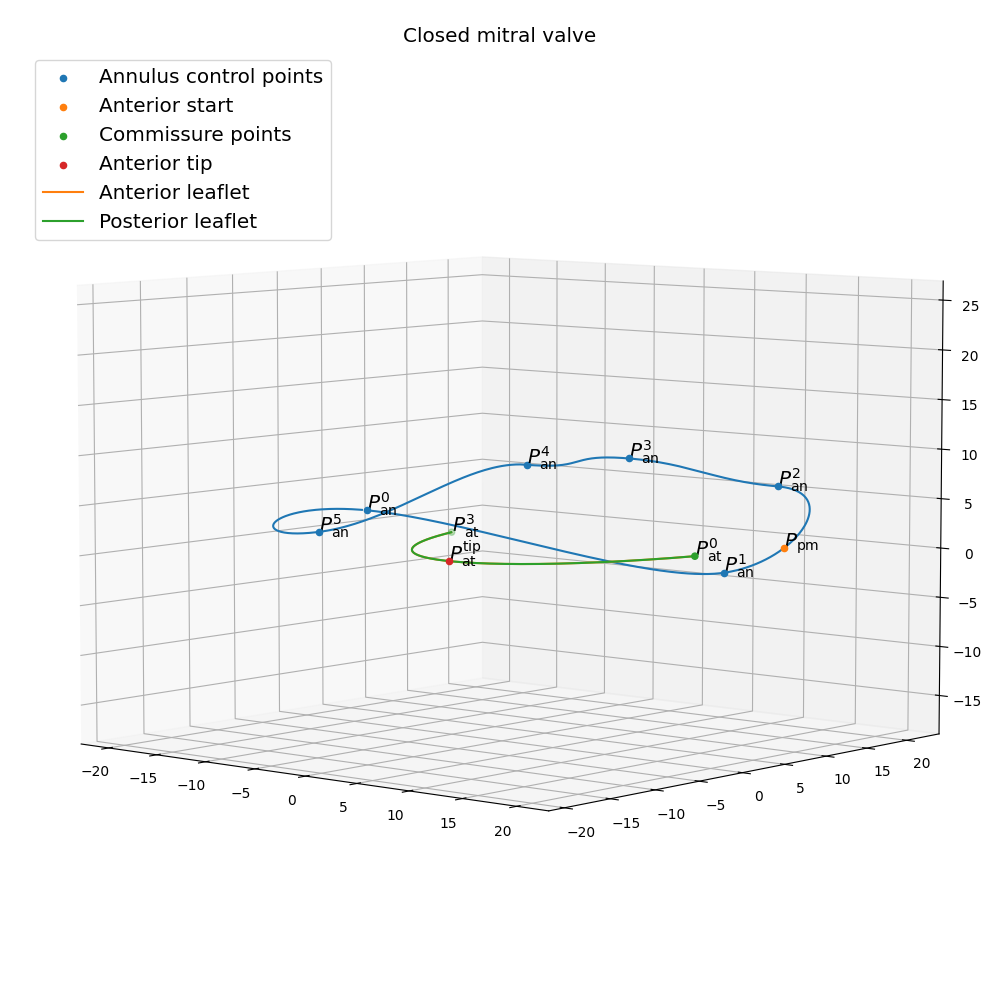}
        \subcaption{Side view}
    \end{subfigure}
    \caption{Top and side view of the mitral valve model in a closed state. Note that
    by construction the anterior and posterior leaflet curves overlap. }
    \label{fig:annulus_closed}
\end{figure}

The annulus' shape is defined by six points $P_{\mathrm{an}}^0$ through
$P_{\mathrm{an}}^5$, which are shown in blue in Figure~\ref{fig:annulus_closed}. The
definitions of the points are
\begin{align*}
    P_{\mathrm{an}}^0 &=
    \begin{pmatrix} 0 \\                 -0.5 \cdot d_{\mathrm{ap}} \\
        h_{\mathrm{po}}
    \end{pmatrix}, \\
    P_{\mathrm{an}}^1 &=
    \begin{pmatrix} 0.5 \cdot d_{\mathrm{alpm}} \\            d_{\mathrm{ap}}  \cdot
        (r_{\mathrm{alpm}} -0.5) \\    0
    \end{pmatrix}, \\
    P_{\mathrm{an}}^2 &=
    \begin{pmatrix} 0.5 \cdot d_{\mathrm{it}} \\  0.5  \cdot (d_{\mathrm{ap}} +
        d_{\mathrm{id}}) \\        0.75  \cdot h_{\mathrm{at}}
    \end{pmatrix}, \\
    P_{\mathrm{an}}^3 &=
    \begin{pmatrix} 0 \\                 0.5 \cdot d_{\mathrm{ap}} \\
        h_{\mathrm{at}}
    \end{pmatrix}, \\
    P_{\mathrm{an}}^4 &=
    \begin{pmatrix} -0.5 \cdot d_{\mathrm{it}} \\ 0.5 \cdot (d_{\mathrm{ap}} +
        d_{\mathrm{id}}) \\         0.75 h_{\mathrm{at}}
    \end{pmatrix}, \\
    P_{\mathrm{an}}^5 &=
    \begin{pmatrix} -d_{\mathrm{it}} \\           d_{\mathrm{ap}} \cdot
        (r_{\mathrm{alpm}} -0.5) \\     0
    \end{pmatrix}.
\end{align*}
These points and the tangents at these points are used to define the control points
for Bézier curves between each neighboring set of points. Each point's tangent
direction is predefined and aligns with either the x or y axis. The tangents'
magnitudes are calculated using fixed ratios of the input parameters, where the ratios
are defined based on empirically found values. The tangents $T_0$ through $T_5$ are defined as
\begin{align*}
    T_{\mathrm{an}}^0 &=
    \begin{pmatrix} 0.5 \cdot 0.55 \cdot (d_{\mathrm{ap}} \cdot r_{\mathrm{alpm}} +
        0.5 \cdot d_{\mathrm{alpm}}) \\ 0 \\ 0
    \end{pmatrix}, \\
    T_{\mathrm{an}}^1 &=
    \begin{pmatrix} 0 \\ 0.55 \cdot (0.5 \cdot d_{\mathrm{id}} + d_{\mathrm{ap}}
        \cdot(1 - r_{\mathrm{alpm}})) \\ 0
    \end{pmatrix}, \\
    T_{\mathrm{an}}^2 &=
    \begin{pmatrix} -0.25 \cdot d_{\mathrm{it}} \\ 0 \\ 0
    \end{pmatrix}, \\
    T_{\mathrm{an}}^3 &=
    \begin{pmatrix} -0.25 \cdot d_{\mathrm{it}} \\ 0 \\ 0
    \end{pmatrix}, \\
    T_{\mathrm{an}}^4 &=
    \begin{pmatrix} -0.25 \cdot d_{\mathrm{it}} \\ 0 \\ 0
    \end{pmatrix}, \\
    T_{\mathrm{an}}^5 &=
    \begin{pmatrix} 0 \\ - 0.55 \cdot (0.5 \cdot d_{\mathrm{id}} + d_{\mathrm{ap}}
        \cdot(1 - r_{\mathrm{alpm}})) \\ 0
    \end{pmatrix}. \\
\end{align*}

Using the endpoints $P_{\mathrm{an}}^i$ and tangents $T_{\mathrm{an}}^i$, the four
control points $R_{\mathrm{an}}^i$ for each Bézier curve are defined as
\begin{align*}
    R_{\mathrm{an}}^i = (P_{\mathrm{an}}^i, P_{\mathrm{an}}^i + T_{\mathrm{an}}^i,
    P_{\mathrm{an}}^{i+1} - T_{\mathrm{an}}^{i+1}, P_{\mathrm{an}}^{i+1}),
\end{align*}
where $i+1$ is calculated modulo 6, such that there is also a Bézier curve from
$P_{\mathrm{an}}^5$ to $P_{\mathrm{an}}^0$. All Bézier curves are combined into a
composite Bézier, which is then resampled such that the composite Bézier is sampled
uniformly. This gives the annulus curve $A(u)$ with $u \in [0, 1]$. Because each
tangent is used for both Bézier curves at a given point $P_{\mathrm{an}}^i$, the
annulus' derivative is continuous and thus the composite Bézier is $C^{1}$ continuous.

When the mitral valve is closed, the anterior and posterior points are higher, giving
the annulus its saddle shape.

\begin{figure}[htbp]
    \centering
    \begin{subfigure}{0.49\linewidth}
        \includegraphics[width=\linewidth]{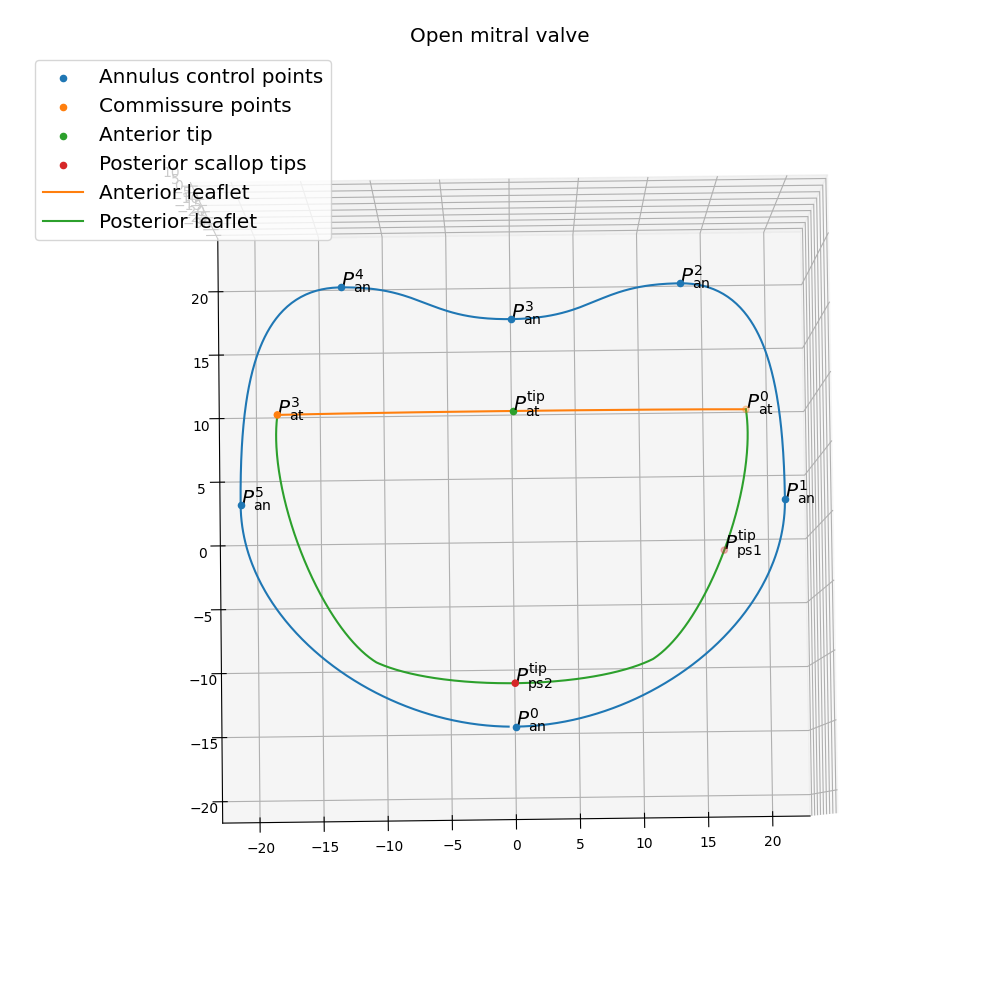}
        \subcaption{Top view}
    \end{subfigure}
    \begin{subfigure}{0.49\linewidth}
        \includegraphics[width=\linewidth]{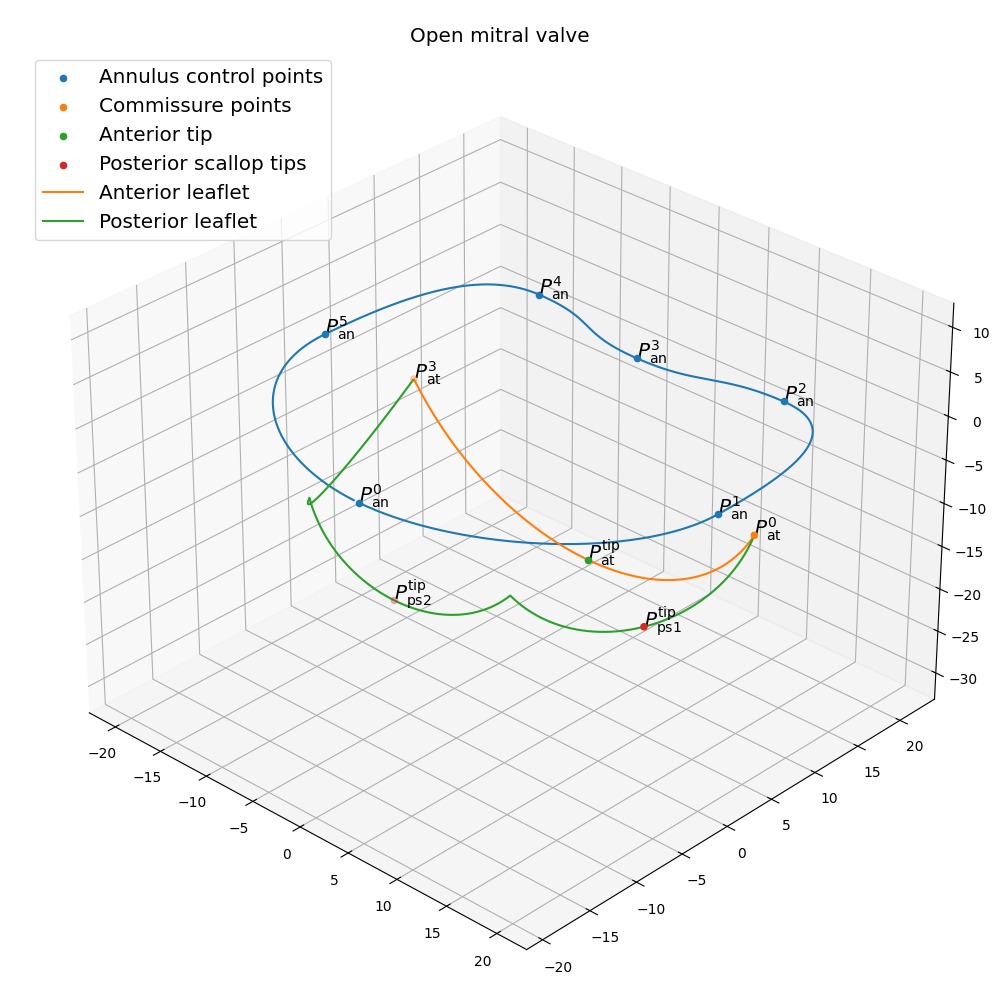}
        \subcaption{Side view}
    \end{subfigure}
    \caption{Top and side view of the mitral valve model in an open state.}
    \label{fig:annulus_open}
\end{figure}

When the mitral valve is open all points are on the xy-plane and the annulus is flat.
This is done by setting the z coordinates to $0$ for all endpoints
$P_{\mathrm{an}}^i$. The open state of the annulus can be seen in
Figure~\ref{fig:annulus_open}. Currently the annulus does not expand when it is in its
open state, though that could be added in the future.

\subsubsection{Leaflets}
The leaflets' curves are calculated using three points, which are the two endpoints
and the tip point. The locations of the endpoints are determined by the commissure
points on the annulus and the commissure width. The locations of the posteromedial and
anterolateral commissures $P_{\mathrm{pm}}$ and $P_{\mathrm{al}}$ on the annulus by
first calculating the relative values $u_{\mathrm{pm}}$ and $u_{\mathrm{al}}$, which
depend on the ratio $r_{\mathrm{at}}$ of the anterior circumference to the total
circumference of the annulus:
\begin{align*}
    u_{\mathrm{pm}} &= 0.5 - \frac{r_{\mathrm{at}}}{2}, \\
    u_{\mathrm{al}} &= 0.5 + \frac{r_{\mathrm{at}}}{2}, \\
    P_{\mathrm{pm}} &= A(u_{\mathrm{pm}}), \\
    P_{\mathrm{al}} &= A(u_{\mathrm{al}}).
\end{align*}
$P_{\mathrm{pm}}$ is shown in Figure~\ref{fig:annulus_closed} in orange. To calculate
the leaflet endpoint positions, we define the annulus inwards 'normal' $N(u)$ for $u
\in [0,1]$ as:
\begin{align}
    N(u) &= \frac{
        \begin{pmatrix} -1 \\ 1 \\ 0
    \end{pmatrix} \odot A'(u)}{\norm{
            \begin{pmatrix} -1 \\ 1 \\ 0
    \end{pmatrix} \odot A'(u)}},
\end{align}
where $\odot$ is defined as element-wise (Hadamard) multiplication of two vectors.
This gives the tangent of $A(u)$ that is projected onto the xy-plane, rotated
counterclockwise by $90^{\circ}$  and normalized.

Instead of setting the control points $P^1$ and $P^2$ for the leaflets directly, they
are calculated indirectly using the leaflet's tip position. This is done by first
approximating a circular arc with angle $\alpha_{\mathrm{a}}$ on the unit circle. This
arc is constructed by setting the control points $S^i$ as described in \cite{farin_nurbs_1999}:
\begin{align*}
    k   &= \frac{4}{3} \cdot \tan \left(\frac{\alpha_{\mathrm{a}}}{4} \right), \\
    S^3 &=
    \begin{pmatrix} \cos(\alpha_{\mathrm{a}}) \\ \sin(\alpha_{\mathrm{a}}) \\ 0
    \end{pmatrix}, \\
    S^2 &= R^3 + k \cdot
    \begin{pmatrix} \sin(\alpha_{\mathrm{a}}) \\ - \cos(\alpha_{\mathrm{a}}) \\ 0
    \end{pmatrix}, \\
    S^1 &=
    \begin{pmatrix} 1 \\ k \\ 0
    \end{pmatrix},    \\
    S^0 &=
    \begin{pmatrix} 1 \\ 0 \\ 0
    \end{pmatrix}.
\end{align*}
Additionally, the tip and center points $S^{\mathrm{tip}}$ and $S^{\mathrm{cen}}$ on
the arc are calculated as

\begin{align*}
    S^{\mathrm{tip}} &=
    \begin{pmatrix} \cos(\frac{\alpha_{\mathrm{a}}}{2})
        \\ \sin(\frac{\alpha_{\mathrm{a}}}{2}) \\ 0
    \end{pmatrix},    \\
    S^{\mathrm{cen}} &= \frac{S^0 + S^3} {2}.
\end{align*}

Using these points, a transformation matrix $M$ is derived that maps $S^0, S^3,
S^{\mathrm{tip}}, S^{\mathrm{cen}}$ to $P^0, P^3, P^{\mathrm{tip}}, P^{\mathrm{cen}}$:
\begin{align*}
    M &=
    \begin{pmatrix}
        & & & \\ 
        S^0 & S^3 & S^{\mathrm{tip}} & S^{\mathrm{cen}} \\
        & & & \\
        1 & 1 & 1 & 1
    \end{pmatrix}
    \begin{pmatrix}
        & & & \\
        P^0 & P^3 & P^{\mathrm{tip}} & P^{\mathrm{cen}}\\
        & & & \\
        1 & 1 & 1 & 1
    \end{pmatrix} ^ {-1}.
\end{align*}

The leaflet control points can then be obtained by multiplying $M$ with the arc control points:
\begin{align}\label{eq:matrix}
    \begin{pmatrix}
        & & & \\
        P^0 & P^1 & P^3 & P^4\\
        & & & \\
        1 & 1 & 1 & 1
    \end{pmatrix}
    &= M
    \begin{pmatrix}
        & & & \\
        S^0 & S^1 & S^3 & S^4\\
        & & & \\
        1 & 1 & 1 & 1
    \end{pmatrix}
\end{align}
Although this method of obtaining the control points for the leaflets is more
complicated than setting them directly, it ensures that the leaflets all follow the
same curvature, which greatly simplifies the model. It also allows changing the
curvature of the leaflets with only a single parameter $\alpha_{\mathrm{a}}$.

\paragraph{Closed valve}
When the leaflets are in the closed position, it is assumed in the model that they
close perfectly along the leaflet edges and thus the anterior and posterior leaflet
curves are shared. 

The leaflet endpoint positions $P_{\mathrm{at}}^0$, $P_{\mathrm{at}}^3$, tip
$P_{\mathrm{at}}^{\mathrm{tip}}$ and center point $P_{\mathrm{at}}^{\mathrm{cen}}$ are
calculated as
\begin{align*}
    P_{\mathrm{at}}^0        &=
    \begin{pmatrix} 1 \\ 1 \\ 0
    \end{pmatrix} \odot P_{\mathrm{pm}} + h_{\mathrm{pm}} \cdot N(u_{\mathrm{pm}}), \\
    P_{\mathrm{at}}^3        &=
    \begin{pmatrix} 1 \\ 1 \\ 0
    \end{pmatrix} \odot P_{\mathrm{al}} + h_{\mathrm{al}} \cdot N(u_{\mathrm{al}}), \\
    P_{\mathrm{at}}^{\mathrm{tip}}    &=
    \begin{pmatrix} 1 \\ 1 \\ 0
    \end{pmatrix} \odot A(0.5) + l_{\mathrm{at}} \cdot N(0.5), \\
    P_{\mathrm{at}}^{\mathrm{cen}}    &= \frac{P^0 + P^3} {2}.
\end{align*}
These are then used to obtain $M$ and also calculate $P_{\mathrm{at}}^1$ and
$P_{\mathrm{at}}^2$ for the leaflet curve. Note that the leaflet curve is on the
xy-plane for simplicity. An example of the leaflets in the closed state is shown in
Figure~\ref{fig:annulus_closed}.

\paragraph{Open valve}
When the mitral valve is in the open state, the leaflets are positioned under the
annulus with an inwards angle $\alpha_{\mathrm{in}}$. The endpoints and tips are
calculated using the function $U(u, x)$ with $u \in [0,1], x \in \mathbb{R}$:
\begin{align*}
    U(u, x)    &= A(u) + x \cdot \frac{
        \begin{pmatrix} \cos(\alpha_{\mathrm{in}}) \\ \cos(\alpha_{\mathrm{in}})
            \\ -\sin(\alpha_{\mathrm{in}})
    \end{pmatrix} \odot N(u)}{\norm{
            \begin{pmatrix} \cos(\alpha_{\mathrm{in}}) \\ \cos(\alpha_{\mathrm{in}})
                \\ -\sin(\alpha_{\mathrm{in}})
    \end{pmatrix} \odot N(u)}}.
\end{align*}
$\alpha_{\mathrm{in}} = 90^\circ$ returns a vector that points down and
$\alpha_{\mathrm{in}} < 90^\circ$ returns a vector that points inwards, similar to a
cone under a circle. The anterior leaflet endpoints $P_{\mathrm{at}}^0$,
$P_{\mathrm{at}}^3$ and tip $P_{\mathrm{at}}^{\mathrm{tip}}$ are defined as
\begin{align*}
    P_{\mathrm{at}}^0 &= U(u_{\mathrm{pm}}, l_{\mathrm{pm}}),   \\
    P_{\mathrm{at}}^3 &= U(u_{\mathrm{al}}, l_{\mathrm{pm}}),   \\
    P_{\mathrm{at}}^{\mathrm{tip}} &= U(0.5, l_{\mathrm{at}}).
\end{align*}
$P_{\mathrm{at}}^1$, $P_{\mathrm{at}}^2$ are then calculated using
Equation~\eqref{eq:matrix} and the anterior leaflet curve is generated using all control points.

The posterior curve in the open state is constructed by combining three cubic Bézier
curves, one for each scallop. The widths of the scallops are all equal, which is one
third of the anterior circumference of the annulus. The length of the middle scallop
$l_{ps2}$ and the lengths of the side scallops $l_{ps1}$ can be set independently. The
scallop endpoints and tips are calculated as
\begin{align*}
    P_{ps1}^0 &= P_{\mathrm{at}}^0 ,\\
    P_{ps1}^3 &= U \left(\frac{u_{\mathrm{pm}}}{3}, l_{\mathrm{ps1}} \right) ,\\
    P_{ps1}^{\mathrm{tip}} &= U \left(\frac{2 \cdot u_{\mathrm{pm}}}{3},
    l_{\mathrm{pm}} \right) ,\\
    P_{ps2}^0 &= P_{\mathrm{ps1}}^3 ,\\
    P_{ps2}^3 &= U \left(1 - \frac{2 \cdot u_{\mathrm{pm}}}{3}, l_{\mathrm{pm}} \right)  ,\\
    P_{ps2}^{\mathrm{tip}} &= U(0, l_{\mathrm{ps2}}) ,\\
    P_{ps3}^0 &= P_{\mathrm{ps2}}^3  ,\\
    P_{ps3}^3 &= P_{\mathrm{at}}^3 ,\\
    P_{ps3}^{\mathrm{tip}} &= U \left(1 - \frac{2 \cdot u_{\mathrm{pm}}}{3},
    l_{\mathrm{pm}} \right) .
\end{align*}
Again, the control points $P^1, P^2$ for each scallop are obtained using
Equation~\eqref{eq:matrix} and the Bézier curve is created from those control points.
The three Bézier curves are then sampled and combined into a single curve, which is
then resampled such that the points are distributed uniformly.

\subsubsection{Leaflet surfaces}\label{sec:mitral_surfaces}
The leaflet surfaces are constructed by drawing quadratic Bézier curves from a point
$A(u)$ on the annulus to $L(u)$ on the leaflets with $s_{sc}$ samples of $u$ and $u
\in [0,1]$. Here $L_{\mathrm{at}}(v)$ and $L_{\mathrm{po}}(v)$ are the anterior and
posterior leaflet curves respectively and $L(u)$ is a mapping of $L_{\mathrm{at}}(v)$
and $L_{\mathrm{po}}(v)$ with $v \in [0,1]$, where the leaflet choice depends on the
value of $u$. $L(u)$ is calculated as:
\begin{align*}
    L(u) &=
    \begin{cases}
        L_{\mathrm{po}}(0.5 - \frac{u}{2 \cdot u_{\mathrm{pm}}})
        ,& \text{if } u < u_{\mathrm{pm}} \\
        L_{\mathrm{at}}(0)
        ,& \text{if } u_{\mathrm{pm}} \leq u \leq u_{\mathrm{pm}} + r_{\mathrm{tr}} \\
        L_{\mathrm{at}}(\frac{u - u_{\mathrm{pm}} - r_{\mathrm{tr}}}{u_{\mathrm{al}} -
        u_{\mathrm{pm}}- 2 \cdot r_{\mathrm{tr}}}) ,& \text{if } u_{\mathrm{pm}} +
        r_{\mathrm{tr}} < u < u_{\mathrm{al}} - r_{\mathrm{tr}} \\
        L_{\mathrm{at}}(1)
        ,& \text{if } u_{\mathrm{al}} - r_{\mathrm{tr}} \leq u \leq u_{\mathrm{al}} \\
        L_{\mathrm{po}}(1 - \frac{u - u_{\mathrm{al}}}{1 - r_{\mathrm{at}}})
        ,& \text{if } u_{\mathrm{al}} < u \\
    \end{cases} .
\end{align*}
Near the commissure points there are triangle regions where a range of points on the
annulus are mapped to the endpoints of the anterior leaflet. This triangle region is
needed to avoid overlapping surfaces near the commissures.

\begin{figure}[htbp]
    \centering
    \begin{subfigure}{0.49\linewidth}
        \includegraphics[width=\linewidth]{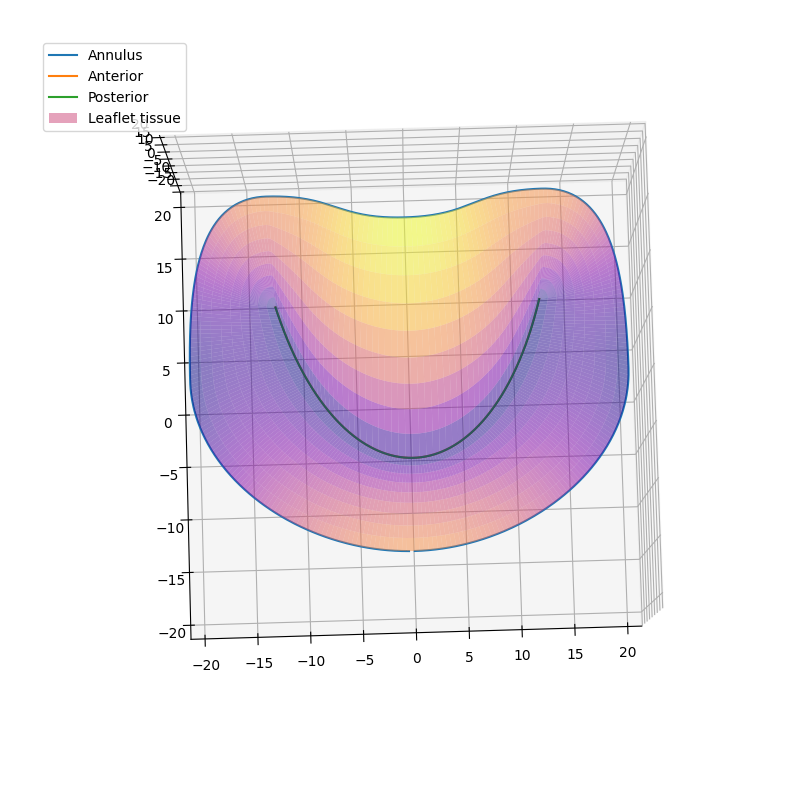}
        \subcaption{Top view}
    \end{subfigure}
    \begin{subfigure}{0.49\linewidth}
        \includegraphics[width=\linewidth]{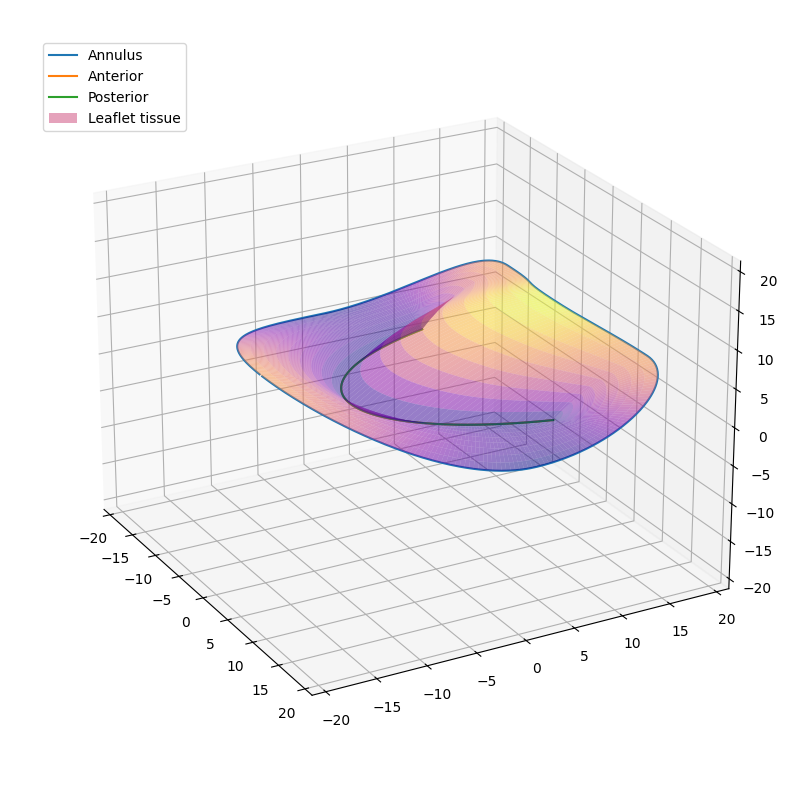}
        \subcaption{Side view}
    \end{subfigure}
    \caption{Closed mitral valve surfaces.}
    \label{fig:surfaces_closed}
\end{figure}

The endpoints of each quadratic surface curve are $A(u)$ and $L(u)$ and the middle
point is determined by the surface curve angle $a_{\mathrm{s}}$, which is the angle
between the surface curve and a straight line from $A(u)$ to $L(u)$. The middle point
$K(u)$ is calculated as:

\begin{align*}
    K(u) &= \frac{A(u) + L(u)}{2} + \tan(\alpha_{\mathrm{s}}) \cdot \frac{L'(u) \times
    A'(u)}{\norm{L'(u) \times A'(u)}} .
\end{align*}
The formula for the quadratic surface curve $D(u, v)$ is then
\begin{align*}
    D(u, v) =&  (1 - v)^2 \cdot A(u) + 2 \cdot (1-v) \cdot v \cdot K(u) \\& {} + v^2 \cdot L(u) ,
\end{align*}
with $v \in [0, 1]$ representing the relative point on the curve at $u$. The surface
itself is then created by drawing quads between each adjacent pair of quadratic
surface curves. An example of the surfaces in a closed and open state are shown in
Figures~\ref{fig:surfaces_closed} and~\ref{fig:mitral_valve} respectively.

A quadratic curve is chosen for the surface instead of a cubic because a quadratic
Bézier is less computationally expensive. The surface is sampled with hundreds of
curves and thus the computational time needed for the surface is decreased
considerably when using quadratic Bézier curves.

\subsection{Results and discussion}

\begin{figure}
    \centering
    \begin{subfigure}{0.4\linewidth}
        \includegraphics[width=\linewidth]{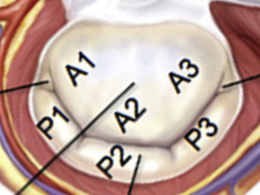}
        \subcaption{Closed reference valve}
    \end{subfigure}
    \begin{subfigure}{0.4\linewidth}
        \includegraphics[width=\linewidth]{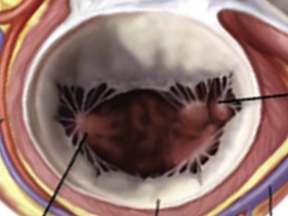}
        \subcaption{Open reference valve}
    \end{subfigure}
    \\
    \begin{subfigure}{0.49\linewidth}
        \includegraphics[width=\linewidth]{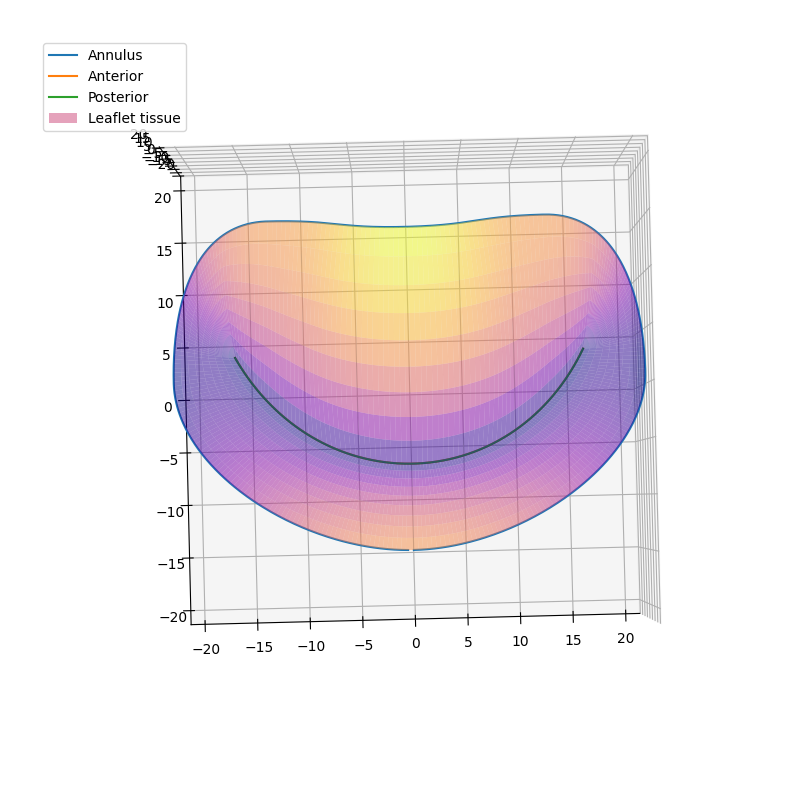}
        \subcaption{Closed valve model}
    \end{subfigure}
    \begin{subfigure}{0.49\linewidth}
        \includegraphics[width=\linewidth]{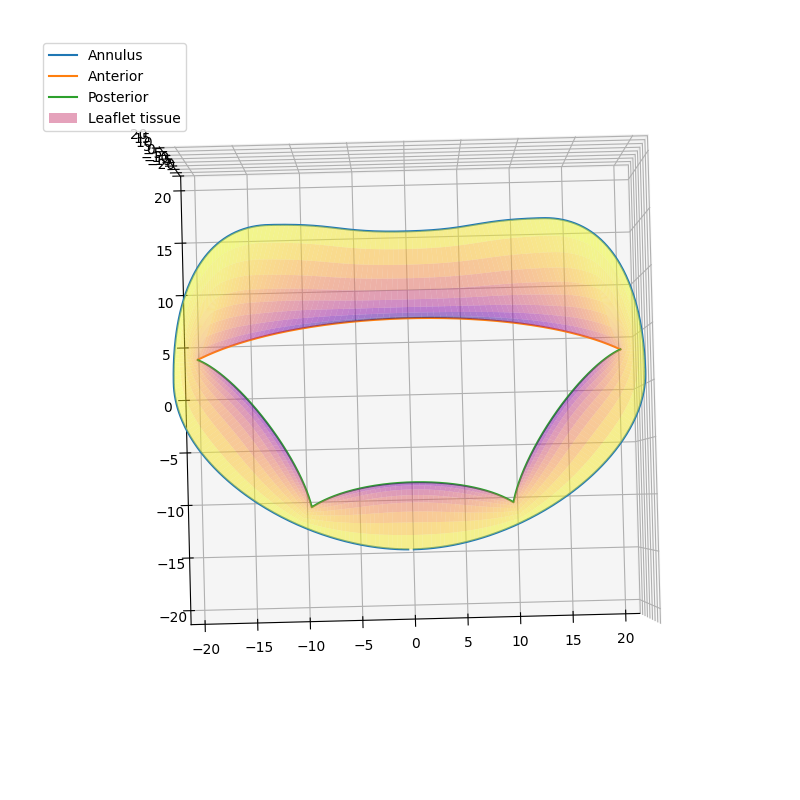}
        \subcaption{Open valve model}
    \end{subfigure}
    \caption{Comparison between the mitral valve model fit to a reference image and
    the image itself. The reference image is zoomed in from Figure~\ref{fig:top_heart}. }
    \label{fig:comparison}
\end{figure}

Figure~\ref{fig:comparison} shows the mitral valve where the parameters are manually
fit to a reference image.
The model is adjusted to match the image by first setting the general shape of the
annulus with $d_{\mathrm{ap}}$, $d_{\mathrm{alpm}}$, $d_{\mathrm{it}}$ and
$d_{\mathrm{id}}$. Unfortunately in this case, the anterior and posterior heights
$h_{\mathrm{at}}$ and $h_{\mathrm{po}}$ cannot be estimated because of the lack of 3D
data. Then, the leaflet curve in closed position is fit to the image by setting
$r_{\mathrm{at}}$, $l_{\mathrm{at}}$, $\alpha_{\mathrm{a}}$ and $h_{\mathrm{pm}}$.
Next, the image of the open mitral valve can be used to determine the values of
$\alpha_{\mathrm{in}}$, $l_{\mathrm{ps1}}$ and $l_{\mathrm{ps2}}$. As a last step all
parameters are tweaked slightly to increase the resemblance of the model with respect
to the image.

Overall, the shape of the model's annulus follows the reference image when closed, but
in the open state the reference annulus becomes more round, whereas the model does
not. The reference also has relatively wide commissures, which are simplified to a
single point in the model. Possible extensions of the model could consider adding a
new commissure. 


\section{Model parameters}

Table~\ref{tab:params_aortic} shows the parameters used for the aortic valve model and
Table~\ref{tab:params_mitral} for the mitral valve model.
\begin{table}
    \begin{subtable}[t]{0.4\linewidth}
        \begin{tabular}[t]{|ll|}
            \hline
            \multicolumn{2}{|c|}{\textbf{Aortic valve parameters}} \\
            \hline
            \textbf{parameter} & \textbf{value} \\
            \multicolumn{2}{|c|}{general} \\
            root\_height        & 1 cm\\
            root\_radius        & 1.29 cm\\
            leaflet\_height& 1.02 cm\\
            bending\_height& 0.71 cm \\
            t3\_leaflet& $[0.5, 0.5, 0.65]$ \\
            t3\_bending& $[0.55, 0.23, 0.4]$ \\
            t0\_leaflet& $[0.025, 0.1, 0.1]$ \\
            t0\_bending& $[0.1, 0.1, 0.1]$ \\
            \hline
            \multicolumn{2}{|c|}{right cusp} \\
            leaflet\_radius\_in & 0.24 cm   \\
            leaflet\_radius\_out & 1.365 cm\\
            leaflet\_angle& $61^{\circ}$    \\
            leaflet\_power& 1.58    \\
            bending\_radius\_in & 0.4 cm           \\
            bending\_radius\_out & 1.39 cm         \\
            bending\_angle& $47.3^{\circ}$    \\
            bending\_power& 1.82          \\
            translation& [0, 0, -1] \\ \hline
            \multicolumn{2}{|c|}{left cusp} \\
            leaflet\_radius\_in& 0.49 cm              \\
            leaflet\_radius\_out& 1.495 cm            \\
            leaflet\_angle& $58.5 ^{\circ}$              \\
            leaflet\_power& 2.77                    \\
            bending\_radius\_in& 0.74 cm             \\
            bending\_radius\_out& 1.355 cm       \\
            bending\_angle& $43 ^{\circ}$                  \\
            bending\_power& 4.12                \\
            rotation& $[0, 0, 119.5]$              \\
            translation& $[0, 0, -1]$  \\ \hline
            \multicolumn{2}{|c|}{non-coronary cusp} \\
            leaflet\_radius\_in& 0.2 cm                \\
            leaflet\_radius\_out& 1.38 cm           \\
            leaflet\_angle& $60.5^{\circ}$                    \\
            leaflet\_power& 1.19                     \\
            bending\_radius\_in& 0.67 cm                 \\
            bending\_radius\_out& 1.435 cm              \\
            bending\_angle& $42.75^{\circ}$                  \\
            bending\_power& 2.54                     \\
            rotation& $[0, 0, -121.5]$             \\
            translation& $[0, 0, -1]$ \\ \hline
        \end{tabular}
        \subcaption{Aortic valve. }
        \label{tab:params_aortic}
    \end{subtable}
    \hfill
    \begin{subtable}[t]{0.4 \linewidth}
        \begin{tabular}[t]{|ll|}
            \hline
            \multicolumn{2}{|c|}{\textbf{Mitral valve parameters}}\\
            \hline
            \textbf{parameter} & \textbf{value} \\
            anterior\_ratio& 0.41                        \\
            AP\_diameter& 3.0 cm                            \\
            ALPM\_diameter& 4.0 cm                      \\
            ALPM\_position\_ratio& 0.55                   \\
            IT\_distance& 2.5 cm                    \\
            indent\_depth& 0.5 cm                 \\
            anterior\_horn\_height& 0.8 cm                  \\
            posterior\_horn\_height& 0.6 cm                \\
            anterior\_length& 2.12 cm                   \\
            posteromedial\_commissure\_height& 0.8 cm        \\
            posterior\_scallop\_length1& 1.35 cm             \\
            posterior\_scallop\_length2& 1.35 cm           \\
            leaflet\_arc\_angle& $120^{\circ}$                  \\
            leaflet\_inward\_angle& $75^{\circ}$           \\
            surface\_angle& $20^{\circ}$                            \\
            triangle\_size\_ratio& 0.1                    \\
            t& 1                                        \\
            translation& $[0, 0, 1]$ \\ \hline
        \end{tabular}
        \subcaption{Mitral valve.}
        \label{tab:params_mitral}
    \end{subtable}
    \caption{Parameters used for the heart valve models. Parameters for the aortic
    valve are on the left and for the mitral valve on the right. }
    \label{tab:valve_params}
\end{table}

\bibliography{cardiomath}

@article{aguayoAnalysisObstaclesImmersed2022,
  title = {Analysis of Obstacles Immersed in Viscous Fluids Using {{Brinkman}}'s Law for Steady {{Stokes}} and {{Navier-Stokes}} Equations},
  author = {Aguayo, Jorge and Lincopi, Hugo Carrillo},
  year = {2022},
  journal = {SIAM Journal on Applied Mathematics},
  volume = {82},
  number = {4},
  pages = {1369--1386},
  doi = {10.1137/20M138569X}
}

@article{anneseSplittingSchemesLagrange2022,
  title = {Splitting Schemes for a {{Lagrange}} Multiplier Formulation of {{FSI}} with Immersed Thin-Walled Structure: Stability and Convergence Analysis},
  author = {Annese, Michele and Fern{\'a}ndez, Miguel A and Gastaldi, Lucia},
  year = {2022},
  journal = {IMA Journal of Numerical Analysis},
  doi = {10.1093/imanum/drac004}
}

@article{astorinoFluidstructureInteractionMultibody2009,
  title = {Fluid-Structure Interaction and Multi-Body Contact. {{Application}} to Aortic Valves},
  author = {Astorino, M. and Gerbeau, J. and Pantz, O. and Traore, K.},
  year = {2009},
  journal = {Computer Methods in Applied Mechanics and Engineering},
  volume = {198},
  number = {45-46},
  pages = {3603--3612},
  publisher = {Elsevier},
  doi = {10.1016/j.cma.2008.09.012}
}

@article{astorinoRobustEfficientValve2012,
  title = {A Robust and Efficient Valve Model Based on Resistive Immersed Surface},
  author = {Astorino, M. and Hamers, J. and Shadden, S. and Gerbeau, J.},
  year = {2012},
  journal = {International Journal for Numerical Methods in Biomedical Engineering},
  volume = {28},
  number = {9},
  pages = {937--959},
  publisher = {{John Wiley and Sons}},
  doi = {10.1002/cnm.2474}
}

@article{burmanMechanicallyConsistentModel2022,
  title = {A Mechanically Consistent Model for Fluid-Structure Interactions with Contact Including Seepage},
  author = {Burman, Erik and Fern{\'a}ndez, Miguel A and Frei, Stefan and Gerosa, Fannie M},
  year = {2022},
  journal = {Computer Methods in Applied Mechanics and Engineering},
  volume = {392},
  pages = {114637},
  doi = {10.1016/j.cma.2022.114637}
}

@article{coffeyModernEpidemiologyHeart2016,
  title = {The Modern Epidemiology of Heart Valve Disease},
  author = {Coffey, Sean and Cairns, Benjamin J and Iung, Bernard},
  year = {2016},
  journal = {Heart},
  volume = {102},
  number = {1},
  pages = {75--85},
  publisher = {{BMJ Publishing Group Ltd and British Cardiovascular Society}},
  doi = {10.1136/heartjnl-2014-307020}
}

@article{dal-bianco_anatomy_2013,
  title = {Anatomy of the {{Mitral Valve Apparatus}}},
  author = {{Dal-Bianco}, Jacob P. and Levine, Robert A.},
  year = {2013},
  month = may,
  journal = {Cardiology Clinics},
  volume = {31},
  number = {2},
  pages = {151--164},
  issn = {07338651},
  doi = {10.1016/j.ccl.2013.03.001},
  urldate = {2024-05-30},
  langid = {english}
}

@book{farin_nurbs_1999,
  title = {{{NURBS}}: From Projective Geometry to Practical Use},
  shorttitle = {{{NURBS}}},
  author = {Farin, Gerald E. and Farin, Gerald E.},
  year = {1999},
  edition = {2nd ed},
  publisher = {A.K. Peters},
  address = {Natick, Mass},
  isbn = {978-1-56881-084-3}
}

@article{fedelePatientspecificAorticValve2017,
  title = {A Patient-Specific Aortic Valve Model Based on Moving Resistive Immersed Implicit Surfaces},
  author = {Fedele, Marco and Faggiano, Elena and Ded{\`e}, Luca and Quarteroni, Alfio},
  year = {2017},
  journal = {Biomechanics and modeling in mechanobiology},
  volume = {16},
  number = {5},
  pages = {1779--1803},
  publisher = {Springer},
  doi = {10.1007/s10237-017-0919-1}
}

@article{fernandezUnfittedMeshSemiimplicit2021,
  title = {An Unfitted Mesh Semi-Implicit Coupling Scheme for Fluid-Structure Interaction with Immersed Solids},
  author = {Fern{\'a}ndez, Miguel A and Gerosa, Fannie M},
  year = {2021},
  journal = {International Journal for Numerical Methods in Engineering},
  volume = {122},
  number = {19},
  pages = {5384--5408},
  publisher = {Wiley Online Library},
  doi = {10.1002/nme.6449}
}

@article{fuchsbergerIncorporationObstaclesFluid2022,
  title = {On the {{Incorporation}} of {{Obstacles}} in a {{Fluid Flow Problem Using}} a {{Navier-Stokes-Brinkman Penalization Approach}}},
  author = {Fuchsberger, Jana and Karabelas, Elias and Aigner, Philipp and Niederer, Steven and Plank, Gernot and Schima, Heinrich and Haase, Gundolf},
  year = {2022},
  journal = {Journal of Computational Science},
  volume = {57},
  pages = {101506},
  publisher = {Elsevier},
  doi = {10.1016/j.jocs.2021.101506}
}

@article{fumagalliImagebasedComputationalHemodynamics2020,
  title = {An Image-Based Computational Hemodynamics Study of the {{Systolic Anterior Motion}} of the Mitral Valve},
  author = {Fumagalli, Ivan and Fedele, Marco and Vergara, Christian and Ded{\`e}, Luca and Ippolito, Sonia and Nicol{\`o}, Francesca and Antona, Carlo and Scrofani, Roberto and Quarteroni, Alfio},
  year = {2020},
  journal = {Computers in Biology and Medicine},
  volume = {123},
  pages = {103922},
  doi = {10.1016/j.compbiomed.2020.103922}
}

@article{fumagalliReduced3D0DFluid2025,
  title = {A Reduced {{3D-0D}} Fluid--Structure Interaction Model of the Aortic Valve That Includes Leaflet Curvature},
  author = {Fumagalli, Ivan and Dede', Luca and Quarteroni, Alfio},
  year = {2025},
  month = aug,
  journal = {Biomechanics and Modeling in Mechanobiology},
  volume = {24},
  number = {4},
  pages = {1169--1189},
  issn = {1617-7940},
  doi = {10.1007/s10237-025-01960-9},
  urldate = {2025-08-02},
  langid = {english}
}

@article{heydenMaterialModelingCardiac2015,
  title = {Material Modeling of Cardiac Valve Tissue: {{Experiments}}, Constitutive Analysis and Numerical Investigations},
  author = {Heyden, S. and Nagler, A. and Bertoglio, C. and Bieler, J. and Gee, M. and Wall, W. and Ortiz, M.},
  year = {2015},
  journal = {Journal of Biomechanics},
  volume = {48},
  pages = {4287--4296},
  doi = {10.1016/j.jbiomech.2015.10.043}
}

@article{kaiserDesignbasedModelAortic2021,
  title = {A Design-Based Model of the Aortic Valve for Fluid-Structure Interaction},
  author = {Kaiser, Alexander D and Shad, Rohan and Hiesinger, William and Marsden, Alison L},
  year = {2021},
  journal = {Biomech Model Mechanobiol},
  volume = {20},
  pages = {2413--2435},
  publisher = {Springer},
  doi = {10.1007/s10237-021-01516-7}
}

@article{laadhariNumericalModelingHeart2016,
  title = {Numerical Modeling of Heart Valves Using Resistive {{Eulerian}} Surfaces},
  author = {Laadhari, Aymen and Quarteroni, Alfio},
  year = {2016},
  journal = {International journal for numerical methods in biomedical engineering},
  volume = {32},
  number = {5},
  pages = {e02743},
  publisher = {Wiley Online Library},
  doi = {10.1002/cnm.2743}
}

@article{mccarthy_anatomy_2010,
  title = {Anatomy of the Mitral Valve: Understanding the Mitral Valve Complex in Mitral Regurgitation},
  shorttitle = {Anatomy of the Mitral Valve},
  author = {McCarthy, K. P. and Ring, L. and Rana, B. S.},
  year = {2010},
  month = dec,
  journal = {European Journal of Echocardiography},
  volume = {11},
  number = {10},
  pages = {i3--i9},
  issn = {1525-2167, 1532-2114},
  doi = {10.1093/ejechocard/jeq153},
  urldate = {2024-07-11},
  langid = {english}
}

@article{oliveira_geometric_2020,
  title = {Geometric Description for the Anatomy of the Mitral Valve: {{A}} Review},
  shorttitle = {Geometric Description for the Anatomy of the Mitral Valve},
  author = {Oliveira, Diana and Srinivasan, Janaki and Espino, Daniel and Buchan, Keith and Dawson, Dana and Shepherd, Duncan},
  year = {2020},
  month = aug,
  journal = {Journal of Anatomy},
  volume = {237},
  number = {2},
  pages = {209--224},
  issn = {0021-8782, 1469-7580},
  doi = {10.1111/joa.13196},
  urldate = {2024-04-29},
  langid = {english}
}

@article{paseParametricGeometryModel2023,
  title = {A Parametric Geometry Model of the Aortic Valve for Subject-Specific Blood Flow Simulations Using a Resistive Approach},
  author = {Pase, Giorgia and Brinkhuis, Emiel and De Vries, Tanja and Kosinka, Ji{\v r}{\'i} and Willems, Tineke and Bertoglio, Crist{\'o}bal},
  year = {2023},
  month = jun,
  journal = {Biomechanics and Modeling in Mechanobiology},
  volume = {22},
  number = {3},
  pages = {987--1002},
  issn = {1617-7940},
  doi = {10.1007/s10237-023-01695-5},
  urldate = {2024-09-18},
  langid = {english}
}

@article{Powell_1964,
  title = {An Efficient Method for Finding the Minimum of a Function of Several Variables without Calculating Derivatives},
  author = {Powell, M. J. D.},
  year = {1964},
  month = feb,
  journal = {The Computer Journal},
  volume = {7},
  number = {2},
  pages = {155--162},
  doi = {10.1093/comjnl/7.2.155},
  langid = {english}
}

@misc{renderkitembree_2025,
  title = {{{RenderKit}}/Embree},
  author = {{RenderKit}},
  publisher = {Intel},
  urldate = {2025-06-05},
  copyright = {Apache-2.0}
}

@article{shen_geometric_2017,
  title = {The Geometric Model of the Human Mitral Valve},
  author = {Shen, Xiaoqin and Wang, Tiantian and Cao, Xiaoshan and Cai, Li},
  editor = {Zhang, Daoqiang},
  year = {2017},
  month = aug,
  journal = {PLOS ONE},
  volume = {12},
  number = {8},
  pages = {e0183362},
  issn = {1932-6203},
  doi = {10.1371/journal.pone.0183362},
  urldate = {2024-04-15},
  langid = {english}
}

@inproceedings{thibaut_use_2019,
  title = {Use of a Parametric Finite-Element Model of the Mitral Valve to Assess Healthy and Pathological Valve Behaviors},
  booktitle = {{{ECCOMAS-MSF}}, 2019, 4th Edition},
  author = {Thibaut, Alleau and Lanquetin, Laurent and Salsac, Anne-Virginie},
  year = {2019},
  month = sep,
  address = {Sarajevo, France},
  hal_id = {hal-02491319},
  hal_version = {v1}
}

@article{thisAugmentedResistiveImmersed2020,
  title = {Augmented {{Resistive Immersed Surfaces}} Valve Model for the Simulation of Cardiac Hemodynamics with Isovolumetric Phases},
  author = {This, Alexandre and {Boilevin-Kayl}, Ludovic and Fern{\'a}ndez, Miguel A and Gerbeau, Jean-Fr{\'e}d{\'e}ric},
  year = {2020},
  journal = {International journal for numerical methods in biomedical engineering},
  volume = {36},
  number = {3},
  pages = {e3223},
  publisher = {Wiley Online Library},
  doi = {://doi.org/10.1002/cnm.3223}
}

\listoffigures

\end{document}